\pdfoutput=1

\documentclass[11pt]{article}

\usepackage[final]{acl}

\usepackage{times}
\usepackage{latexsym}

\usepackage[T1]{fontenc}

\usepackage[utf8]{inputenc}

\usepackage{microtype}

\usepackage{inconsolata}

\usepackage{graphicx}
\usepackage{listings}
\usepackage{tcolorbox}
\usepackage{algorithm}
\usepackage{algpseudocode}
\usepackage{float}
\usepackage{fancyvrb}
\usepackage{multirow}
\usepackage{amsmath}
\usepackage{amsfonts}
\usepackage{enumitem}

\title{Quality Assessment of Tabular Data using Large Language Models and Code Generation}

\author{Ashlesha Akella\\
  IBM Research, India\\
  \texttt{ashlesha.akella@ibm.com}
  \\\And
  Akshar Kaul\\
  IBM Research, India\\
  \texttt{akshar.kaul@in.ibm.com}
  \\\AND
  Krishnasuri Narayanam\\
  IBM Research, India\\
  \texttt{knaraya3@in.ibm.com}
  \\\And
  Sameep Mehta\\
  IBM Research, India\\
  \texttt{sameepmehta@in.ibm.com}
  }

\begin{document}
\maketitle

\begin{abstract}

Reliable data quality is crucial for downstream analysis of tabular datasets, yet rule-based validation often struggles with inefficiency, human intervention, and high computational costs. We present a three-stage framework that combines statistical inliner detection with LLM-driven rule and code generation. After filtering data samples through traditional clustering, we iteratively prompt LLMs to produce semantically valid quality rules and synthesize their executable validators through code-generating LLMs. To generate reliable quality rules, we aid LLMs with retrieval-augmented generation (RAG) by leveraging external knowledge sources and domain-specific few-shot examples. Robust guardrails ensure the accuracy and consistency of both rules and code snippets. Extensive evaluations on benchmark datasets confirm the effectiveness of our approach.


\end{abstract}
\section{Introduction}

Data quality (DQ) is vital for business decisions; poor data quality costs organizations an average of $\$12.9$ million annually \cite{gartner}, underscoring the need for rigorous DQ management. Data errors stem from sensor faults, entry mistakes, and poor data integration, producing inconsistencies across today’s high-dimensional tabular datasets from diverse domains, which often contain millions of rows and numerous columns. 

\begin{figure*}[ht]
    \centering
    \includegraphics[width=0.85\linewidth]{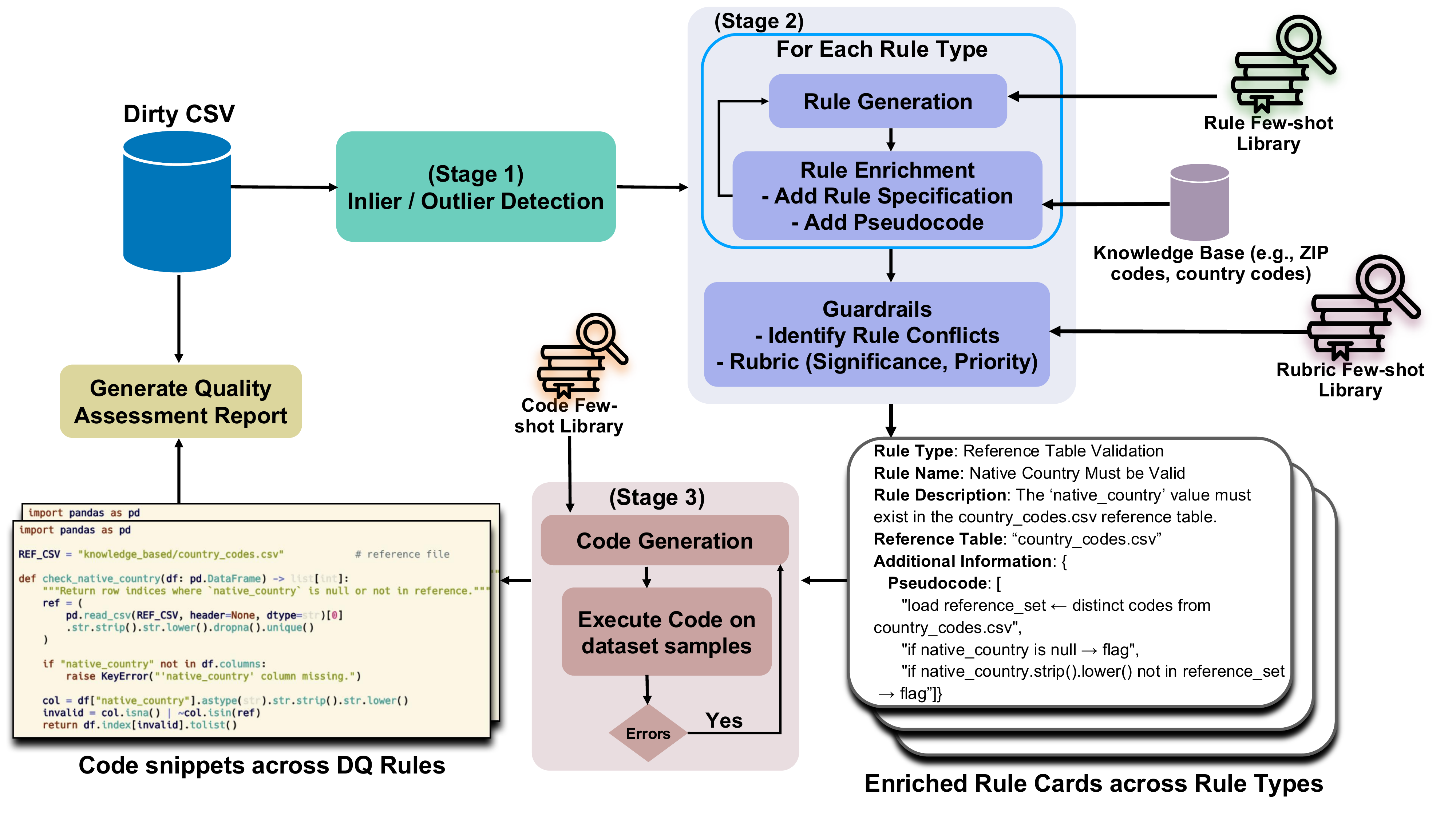}
    \caption{System architecture illustrating the end-to-end pipeline for data quality assessment.}
    \label{fig:arch}
\end{figure*}

Statistical profiling—encompassing distribution shifts, outliers, and functional dependency (FD) violations—remains a foundational technique for detecting data quality issues \cite{bohannon2006conditional,fan2010discovering,krishnan2016activeclean, geerts2020cleaning,rezig2021horizon, jianbin_qin_7a4233c3,toon_boeckling_a196dedc}. Bayesian extensions enhance this approach by modelling expected cell-value posteriors \cite{azzalini2023enhancing, qin2024clean}. While such methods effectively flag structural anomalies, they often lack the semantic understanding required to detect context-dependent errors or violations that rely on external knowledge, leading to overlooked or misclassified issues.

Deep learning approaches learn latent space representations for data cleaning \cite{sigmod19.HoloDetect,jager2024data,reis2024generalizable} or train on constraint-compliant subsets to boost accuracy \cite{biessmann2018deep,nasfi2025improving}.
\citet{vldb21.Picket.selfsupervised} explore self-supervised learning, 
while \citet{reis2024generalizable} propose an active learning-based framework to improve DQ. These methods, however, presuppose clean labels or stable constraints, struggle on heavily noisy tables, and become costly on large datasets. Rule-based data cleaning approaches \cite{vldb07.cfd,vldb08.DQrules,boeckling2022cleaning,boeckling2022efficient} outperform purely statistical techniques, but fail to capture semantic inconsistencies.

Large Language Models (LLMs) offer significant promise for data-quality tasks due to their ability to assess contextual correctness and identify anomalies that traditional FD rules often miss. Yet, existing LLM-based solutions—such as fine-tuned models, prompt-driven detectors \cite{sigmod19.HoloDetect,mehra2024leveraging,zhang2024jellyfish}, or in-context repair mechanisms \cite{DQmetric.VLDB17,LLMclean.ADBIS24,DataRepair.VLDB24,IterClean.ACM.TURC24}—remain computationally expensive, either requiring per-row inference or dataset-specific training. \citet{bendinelli2025exploring} combine an LLM with Python to address cell-level and row-level data quality issues; their approach requires strong hints on the errors in the dataset and fails to address data errors dependent on external domain knowledge. Therefore, an effective solution must balance LLM adaptability while scaling efficiently and incorporate domain-specific knowledge to detect semantically complex errors.

Data quality is typically organized into dimensions such as accuracy, completeness, conformity, and consistency \cite{DQdimensions.MIT96,10.1145/505248.506010,loshin2010practitioner,carlo2018data,fdata22.DQSurvey}, as these categories reflect how researchers and practitioners diagnose defects and prioritize remediation. Building on this taxonomy, we define targeted rule types (see Appendix \ref{sec:dq_rule_background}) under each dimension and generate rules accordingly (see Table \ref{tab:dq_rules}). This design enables granular data quality reports, precise and unambiguous LLM prompts, and modular downstream filters and evaluation metrics.

\begin{table*}[ht!]
\centering
\small
\resizebox{0.97\textwidth}{!}{%
\begin{tabular}{||p{2.4cm}||p{4.4cm}|p{8.5cm}||}
\hline
\textbf{DQ Dimension}        & \textbf{DQ Rule Type}                 & \textbf{Sample DQ Rules (Rule Cards)}  \\
\hline
\multirow{1}{*}{Accuracy}  & Reference Table Verification & 
$\forall x \in \texttt{state}, 
\exists r \in \texttt{Cities} \quad \textit{such that} \quad x = r.state$
\\
\hline
\multirow{3}{*}{Conformity}  & Format Compliance                     & 
$\forall x \in \texttt{beer\_name},  \quad \neg isNull(x) \land \text{ } \texttt{\^{}[A-Za-z0-9' ]+}$ \\ \cline{2-3}
                             & \multirow{2}{*}{Data Type Validation}                  & 
$\forall x \in \texttt{Single Epithelial},\quad \neg isNull(x) \land x \in \mathbb{Z}
$ \\ \cline{3-3}
& &
$\forall x \in \texttt{Marginal Adhesion},\quad \neg isNull(x) \land x \in \mathbb{Z}
$ \\
\hline
\multirow{2}{*}{Completeness}  & \multirow{2}{*}{Missing Value Identification} & 
$\forall x \in \texttt{Clump Thickness}, \neg isNull(x)$
\\ \cline{3-3} 
                             &                                       &
$\forall x \in \texttt{Bare Nuclei},\neg isNull(x)$
\\
\hline
\multirow{3}{*}{Consistency}                    & \multirow{2}{*}{Value Set Constraint}    & $\forall x \in \texttt{Uniformity of Cell Shape}, \neg isNull(x) \ \land \ x \in \{\text{1, 2, 3, 4, 5, 6, 7, 8, 9, 10}\}$
\\ \cline{2-3} 
                                                & \multirow{1}{*}{Cross-Column Validation} &
$\forall r \in \texttt{dataset},\texttt{NormalNucleoli}(r)<\texttt{Mitoses}(r)$
\\
\hline
\end{tabular}}
\caption{Examples of DQ Rules categorized by different DQ Dimensions (more Rules in Table \ref{tab:add_dq_rules} of Appendix). Reference Table Verification and Format Compliance are using Beers while rest using Breast Cancer dataset.}
\label{tab:dq_rules}
\end{table*}

\section{Framework}

Our framework shown in Figure \ref{fig:arch} breaks down the data quality assessment into three stages. First, it uses statistical analysis on the given tabular dataset to categorize each row as either an inlier or an outlier. Then, we perform off-the-shelf LLM inference via automated prompts to generate data quality rules tailored to the dataset. Finally, we generate executable code for each quality rule through an off-the-shelf code-generating LLM inference. We engage only the inlier dataset from the first stage in subsequent stages to ensure reliable analysis utilizing data that is less likely to have errors or inconsistencies.
\subsection{Inlier–Outlier Detection}
\label{sec:inlier-outlier-detection}

Our system initiates preprocessing to identify a subset of non-noisy rows from the input dataset using a traditional clustering technique, executed efficiently through distributed processing on Apache Spark. The pipeline performs multivariate outlier detection at the row level using the Sparx algorithm \cite{Sparx.KDD22}, which scales linearly with the dataset size. For rows flagged as outliers, a finer-grained analysis determines whether specific cells are true outliers or are statistically influenced by other cells, based on the profile of each column. String columns are embedded using BERT \cite{naacl19.bert} and evaluated with a univariate distance-based outlier test, while numeric and categorical columns are analyzed using detectors tailored to their respective data types. A row is ultimately marked as an outlier if the number of its outlier cells surpasses a predefined threshold. This preprocessing step significantly enhances data quality by retaining only the inlier subset, thereby enabling more accurate rule generation and code synthesis in later stages.

\subsection{Generation and Enrichment of DQ Rules}
\label{sec:rule-generation-and-enrichment}

Our system appropriately prompts the Gemma-3-12B \cite{team2025gemma} LLM to elicit DQ Rules for each {\em Rule Type} from the given dataset. For each Rule, Gemma initially formulates a canonical (draft) rule description, followed by a rule enrichment phase to incorporate essential specification and a concise set of pseudocode clauses that precisely capture the validation logic.

We capture each DQ Rule as a \textit{Rule Card}—a structured JSON object comprising of fields, such as: \emph{Rule Name} (a concise title), \emph{Rule Description}, \emph{Target Columns} (the columns involved in the rule), \emph{Specification}, and \emph{Pseudocode}. Organizing rules in this structured format simplifies data quality evaluation by clearly defining the relevant columns and the specific validation objective. Figure \ref{fig:draft_and_enriched_rules_beers_dataset} illustrates a draft and enriched Rule Card.

\begin{figure}[h]
    \tiny
    \setlength{\topsep}{0pt}
    \setlength{\partopsep}{0pt}
    \setlength{\parskip}{0pt}
    \centering
    \begin{tcolorbox}[
        colback=cyan!15!white,
        colframe=cyan!55!white,
        title=\textcolor{black}{\textbf{Rule Card (draft):}},
        boxsep=1pt,
        left=2pt,
        right=2pt,
        top=2pt,
        bottom=1pt,
        width=\linewidth
    ]
    \begin{verbatim}
{
  "Rule Type": "Reference Table Verification",
  "Rule Name": "State Must Follow US State Code Format",
  "Rule Description": "The `state` column must be a two-letter code 
  corresponding to valid US state abbreviations (e.g., NY, CO, CA, FL). 
  Any non-standard two-letter combinations should be flagged as invalid.",
  "Target Columns": ["state"],
  "Reference Table": ["uscities.csv", "Country_phone_codes.csv"]
}
    \end{verbatim}
    \end{tcolorbox}
    \vspace{-5mm}
    \begin{tcolorbox}[
        colback=cyan!15!white,
        colframe=cyan!55!white,
        title=\textcolor{black}{\textbf{Rule Card (enriched):}},
        boxsep=1pt,
        left=2pt,
        right=2pt,
        top=2pt,
        bottom=0pt,
        width=\linewidth
    ]
    \begin{verbatim}
{
  "Rule Type": "Reference Table Verification",
  "Rule Name": "State Must Follow US State Code Format",
  "Rule Description": "The `state` column must contain a two-letter 
  abbreviation (e.g., NY, CO, CA, FL). Any value not on the official list 
  is invalid.",
  "Target Columns": ["state"],
  "Reference Table": "uscities.csv",
  "Additional Information": {
    "Specification": "Validate against the two-letter state_id field in 
`uscities.csv`; ignore `Country_phone_codes.csv`, which is unrelated.",
    "Pseudocode": ["if state is null → flag",
      "if len(state) != 2 → flag",
      "if state.upper() not in us_state_abbrevs_from_csv → flag"]
    } 
  }
    \end{verbatim}
    \end{tcolorbox}
 
    \caption{Draft and Enriched Rule Cards by Gemma-3-12B on Beers dataset 
    \cite{Beers.dataset}}
    \label{fig:draft_and_enriched_rules_beers_dataset}
\end{figure}

\textbf{Pipeline to generate Rules.} It begins by prompting an LLM to generate a schema description (shown in Figure \ref{fig:schema_example} inspired from \cite{zhang2025autoddg}) to incorporate into rule-type-specific prompt templates (refer to Figure \ref{fig:rule_prompt_format_compliance} in Appendix) to guide rule generation.
 

Each rule-generation prompt follows a structured, three-part format. The first part is the \texttt{task header}, which specifies the target rule type (e.g., \textit{Format Compliance} or \textit{Cross-Column Validation}) and includes a detailed description of its intent. The second part comprises \texttt{task blocks} that incorporate contextual elements such as domain-specific examples, previously generated rules (if available), and a knowledge section along with an enumeration of allowed and disallowed behaviors. The third part provides the \texttt{table schema}. For wide tables, the schema is split into manageable batches to fit within Gemma’s context window while still preserving column-level details.


\textbf{Domain-aware few-shot examples.} We maintain a repository of rule-card examples, either handcrafted or harvested from diverse application domains. At run time we embed the table description and every stored domain descriptor with the \textit{all-MiniLM-L6-v2} model \cite{wang2020minilm}. Cosine similarity selects the nearest domain; its representative rule cards for each rule type are then added to the prompt. This domain-specific few-shot context steers Gemma to generate rules with vocabulary and constraint patterns aligned with the target dataset.

\textbf{Few-shot examples from previous iterations.} We employ an iterative prompting strategy inspired by the self-consistency technique in CoT \cite{irlc23.selfConsistentCoT} reasoning. During the first iteration, the LLM is provided with only the schema fragment and a few domain-specific few-shot examples, from which it generates a first batch of rule cards. In subsequent iterations, a randomly selected subset of these generated cards is included as additional exemplars in the prompt, progressively refining the model output.

\textbf{Pipeline to enrich Rules.} A rule-type-agnostic enrichment further refines each generated rule card to enhance specificity and reliability. The enrichment prompt is composed of components, such as: (i) \texttt{Column Profile}: A JSON summary automatically derived from the dataset, detailed in Figure \ref{fig:col_summary}. These column statistics provide the model with concrete and data-driven context. (ii) \texttt{Draft Rule Card}: The preliminary version of the rule generated in the earlier phase. It includes a tentative rule name, description, and target columns—describing the constraint for the model to refine. (iii) \texttt{Clean vs. Noise Sample}: Two short lists of values extracted from the same column—one from inlier (clean) rows and the other from outlier (noisy) rows. Presenting this contrast helps the model recognize practical differences between valid and erroneous data. These components are integrated into the rule enrichment prompt template (see Figure \ref{fig:enrich_rule_prompt_template} in Appendix) and submitted to Gemma-3-12B.


\subsubsection{Multi-Layer Gaurdrails.}
Before the code generation stage, the provisional set of enriched DQ rules passes through a series of validation filters designed to eliminate redundancy, logical inconsistencies, and low-value constraints.

\textbf{Conflict-Resolution Filter.} Early pilot evaluations revealed that the LLM occasionally produced conflicting or mutually exclusive rules. These contradictions stemmed primarily from two sources: (i) inconsistencies in the sample data frames used across iterations; (ii) LLMs hallucinations (see Appendix \ref{sec:examples-of-conflicting-rules} for examples).
To resolve this, the draft rule set specific to `Target  Columns' are passed to an LLM (Gemma) using a dedicated conflict-detection prompt for semantic parsing (see Figure \ref{fig:conflict_resolution_prompt} in Appendix). The model returns a structured JSON report consisting of conflicting rule groups, overlapping target columns, a concise explanation of the conflict, and a recommendation on which rules to discard. Rules flagged for removal are purged before any rubric scoring or downstream validation.
While ours is a rule-type-agnostic conflict detector, there exist logic-programming solutions \cite{ecai20.ruleconflicts} limited to conflict detection among FD rules (e.g., Cross-Column Dependency or Dependency Constraints) but they do not generalize to rules of arbitrary rule types. 

\textbf{Rubric-Based Rule Evaluation} 
An eight-point rubric \cite{rubric-acl24} assesses each surviving rule card. Using a small data sample and the complete table schema, Gemma-3-12B assigns one of these labels to each rule: \textit{Duplicate} (identical to another rule), \textit{Redundant} (subsumed by a stricter rule), \textit{Trivial} (enforced already by schema or data types), \textit{Risk-false-positive} (likely to break as valid data evolves), \textit{Miscategorized} (tagged under the wrong DQ category), \textit{Ambiguous} (unclear wording or missing logic), \textit{Hallucinated-overly-specific} (overly narrow, often unrealistic constraints; e.g., Latitude must be exactly between 40.71271 and 40.71279), \textit{High-value}, \textit{Correct}, or \textit{Correct-fixable}. Only rules categorized as \textit{High-value}, \textit{Correct}, or \textit{Correct-fixable} proceed to the next deterministic filtering phase (prompt in Figure \ref{fig:rubric_prompt}).


\subsection{Python Code Generation}
In the final stage, each rule card is translated into executable Python code using \textit{Qwen-2.5-Coder (32 B)} \cite{bai2023qwen}, an LLM specialized for code generation. This translation enables fine-grained, cell-level validation by producing a Boolean error mask when the generated code is run on the dataset, allowing comparisons against ground truth labels.

To perform this translation, the system constructs a comprehensive prompt for each rule card, integrating five key context components: (i) table schema, (ii) column-summary block containing profiles for all attributes listed in the rule's \texttt{Target Columns}, (iii) the enriched rule card including both specification and pseudocode, (iv) a representative sample of rows from the dataset, and (v) few-shot examples drawn from a curated library of domain- and rule-type–specific code snippets. We dynamically select these few-shot examples by embedding both the current table description and all stored code snippets using sentence embeddings, ranking them by cosine similarity, and inserting the most relevant samples into the prompt. The fully assembled prompt is submitted to the LLM, which generates \textbf{check(df)} function to find invalid cells based on the specified rule. We run each snippet on a few rows; if it errors, the exception is fed back and the model repairs the code (see Figure \ref{fig:code_for_beer_dataset}).

\begin{figure}[ht!]
    \tiny
    \setlength{\topsep}{0pt}
    \setlength{\partopsep}{0pt}
    \setlength{\parskip}{0pt}
    \centering
    \begin{tcolorbox}[
        colback=blue!3!white,
        colframe=blue!45!white,
        title=\textcolor{black}{Code Snippet},
        boxsep=1pt,
        left=2pt,
        right=2pt,
        top=2pt,
        bottom=0pt,
        width=\linewidth
    ]
        \begin{verbatim}
import pandas as pd
REF_FILE = "external_knowledge_base/uscities.csv" 

def check(df: pd.DataFrame) -> list[int]:
    """
    Flag rows whose `state` value is null, not two characters, or not found
    in the official two-letter USPS list from uscities.csv.
    """
    # Load reference abbreviations (uppercase for exact match)
    ref_states = (
        pd.read_csv(REF_FILE, usecols=["state_id"], dtype=str)["state_id"]
        .str.strip().str.upper().unique()
    )
    # Normalise the target column
    col = df["state"].astype(str).str.strip().str.upper()
    # Build invalid mask
    invalid_mask = (
        col.isna() |                  # null
        (col.str.len() != 2) |        # not two letters
        (~col.isin(ref_states))       # not in reference list
    )
    # Return offending row indices
    return sorted(df.index[invalid_mask].tolist())
    \end{verbatim}
    \end{tcolorbox}
    \caption{Executable code snippet generated by Qwen-2.5-Coder on Beers dataset \cite{Beers.dataset}.}
    \vspace{-2mm}
    \label{fig:code_for_beer_dataset}
\end{figure}

\subsection{End-to-End DQ Assessment Workflow}

The framework accepts a single noisy CSV file and sequentially executes each \texttt{check(df)} function generated earlier. Each function returns the row indices that violate its \textit{Rule Card}'s target column. We aggregate these indices into a unified error mask that captures all identified invalid cells across the dataset. Leveraging this mask, the system generates a comprehensive \emph{Quality Assessment Report} that enumerates every flagged cell, along with the name of the triggering rule and the exact Python snippet responsible for its detection to help data stewards in data inspection (see Appendix \ref{sec:Appendix_end_to_end_pipeline} example). 

\section{Experiments}

Our assessment begins with clean benchmark tables from the ED2 \cite{cikm19.ED2} and RAHA \cite{mahdavi2019raha} studies, as well as popular public datasets, into which we systematically inject synthetic errors using the standardized REIN \cite{REIN.edbt23} corruption model. This unified corruption strategy ensures consistency across datasets, allowing fair comparisons with baseline methods widely used in prior work. Then we compare the performance of our framework with baseline approaches under various noise conditions. Subsequently, we assess the contribution of the inlier detection module, evaluate the impact of incorporating domain-specific few-shot examples, and benchmark our approach against established error detection frameworks.


\subsection{Benchmark clean data with unified noise}
\label{sec:benchmark-clean-data-with-unified-noise}


We begin with \emph{clean} versions of ten well-known tabular datasets spanning both transactional and sensor-like sources—namely \textit{Adult} \cite{Adult.dataset}, \textit{Beers} \cite{Beers.dataset}, \textit{Bikes} \cite{Bikes.dataset}, \textit{Breast Cancer} \cite{BreastCancer.dataset}, \textit{HAR} \cite{HAR.dataset}, \textit{Movies} \cite{Movies.dataset}, \textit{Nasa} \cite{Nasa.dataset}, \textit{Rayyan} \cite{ouzzani2016rayyan}, \textit{Soil Moisture} \cite{SoilMoisture.dataset}, and \textit{Tax} \cite{Raha.Benchmark.Datasets}.



To create a balanced yet challenging testbed, we apply the \textbf{five-component error-injection} framework from REIN to each dataset. It introduces diverse error types: \textit{keyboard-based typos}, \textit{explicit and implicit missing values}, \textit{cell swaps}, and \textit{Gaussian noise}. Each error type is injected independently at three noise levels—$10\%$, $20\%$, and $30\%$ of cells—resulting in $30$ corrupted datasets. A set of baseline error detection techniques—\textsc{ED2}, \textsc{FAHES}, \textsc{KATARA} \cite{vldb15.katara}, \textsc{Outlier IQR} \cite{zhang2013advancements}, \textsc{Outlier IF} \cite{liu2012isolation}, \textsc{Outlier SD} \cite{zhang2013advancements}, \textsc{Max-Entropy} \cite{vldb16.maxentropy}, and \textsc{Min-k} \cite{vldb16.maxentropy}—alongside our proposed approach execute on all the corrupted datasets. Table \ref{tab:unified_noise_10_20_30} captures the reported F\textsubscript{1} scores (See Appendix \ref{sec:precision_recall_scores}).

\begin{table*}[ht]
\centering
\small
\resizebox{0.7\textwidth}{!}{%
\begin{tabular}{|c|c|c|c|c|c|c|c|c|c|c|}
\hline
\multicolumn{2}{|c|}{\textbf{Dataset / Noise}} & ED2 & FAHES & KATARA & IQR & IF & SD & Max Entropy & Min-K & Ours \\
\hline\hline
\multirow{3}{*}{Adult}
  & 10\% & \textbf{0.96} & 0.01 & 0.03 & 0.00 & 0.00 & 0.00 & 0.92 & 0.05 & 0.94 \\
  & 20\% & 0.56 & 0.01 & 0.01 & 0.00 & 0.00 & 0.00 & 0.59 & 0.00 & \textbf{0.80} \\
  & 30\% & 0.36 & 0.02 & 0.06 & 0.00 & 0.00 & 0.00 & 0.00 & 0.02 & \textbf{0.63} \\ \hline
\multirow{3}{*}{Beers}
  & 10\% & 0.64 & 0.05 & 0.59 & 0.00 & 0.00 & 0.00 & 0.61 & 0.61 & \textbf{0.72} \\
  & 20\% & 0.80 & 0.04 & 0.52 & 0.00 & 0.00 & 0.00 & 0.75 & 0.55 & \textbf{0.82} \\
  & 30\% & 0.71 & 0.03 & 0.52 & 0.00 & 0.00 & 0.00 & 0.53 & 0.53 & \textbf{0.85} \\ \hline
\multirow{3}{*}{Bikes}
  & 10\% & 0.16 & 0.01 & 0.13 & 0.00 & 0.00 & 0.00 & 0.00 & 0.01 & \textbf{0.82} \\
  & 20\% & 0.39 & 0.01 & 0.23 & 0.00 & 0.00 & 0.00 & 0.00 & 0.00 & \textbf{0.83} \\
  & 30\% & 0.39 & 0.02 & 0.24 & 0.00 & 0.00 & 0.00 & 0.00 & 0.10 & \textbf{0.82} \\ \hline
\multirow{3}{*}{Breast Cancer}
  & 10\% & 0.13 & 0.01 & 0.14 & 0.00 & 0.00 & 0.00 & 0.00 & 0.00 & \textbf{0.90} \\
  & 20\% & 0.46 & 0.09 & 0.09 & 0.00 & 0.06 & 0.00 & 0.36 & 0.19 & \textbf{0.74} \\
  & 30\% & 0.34 & 0.04 & 0.32 & 0.00 & 0.00 & 0.00 & 0.33 & 0.41 & \textbf{0.89} \\ \hline
\multirow{3}{*}{HAR}
  & 10\% & 0.32 & 0.00 & 0.07 & 0.00 & 0.00 & 0.00 & 0.00 & 0.00 & \textbf{0.42} \\
  & 20\% & 0.53 & 0.00 & 0.05 & 0.10 & 0.00 & 0.00 & 0.10 & 0.07 & \textbf{0.54} \\
  & 30\% & 0.49 & 0.00 & 0.12 & 0.00 & 0.00 & 0.00 & 0.31 & 0.00 & \textbf{0.55} \\ \hline
\multirow{3}{*}{Movies}
  & 10\% & \textbf{0.75} & 0.07 & 0.01 & 0.00 & 0.00 & 0.00 & 0.00 & 0.00 & 0.71 \\
  & 20\% & \textbf{0.76} & 0.11 & 0.02 & 0.00 & 0.00 & 0.00 & 0.54 & 0.32 & 0.71 \\
  & 30\% & \textbf{0.77} & 0.07 & 0.03 & 0.00 & 0.00 & 0.00 & 0.00 & 0.00 & 0.68 \\ \hline
\multirow{3}{*}{Nasa}
  & 10\% & 0.88 & 0.03 & 0.00 & 0.00 & 0.00 & 0.00 & 0.80 & 0.05 & \textbf{0.89} \\
  & 20\% & 0.74 & 0.14 & 0.00 & 0.00 & 0.00 & 0.00 & 0.65 & 0.14 & \textbf{0.85} \\
  & 30\% & 0.91 & 0.05 & 0.00 & 0.00 & 0.00 & 0.00 & \textbf{0.93} & 0.05 & 0.91 \\
  \hline
\multirow{3}{*}{Rayyan}
  & 10\% & \textbf{0.94} & 0.16 & 0.35 & 0.00 & 0.00 & 0.00 & \textbf{0.94} & 0.42 & \textbf{0.94} \\
  & 20\% & 0.95 & 0.22 & 0.35 & 0.00 & 0.00 & 0.00 & 0.95 & 0.47 & \textbf{0.97} \\
  & 30\% & 0.95 & 0.17 & 0.35 & 0.00 & 0.00 & 0.00 & 0.95 & 0.47 & \textbf{0.96} \\ \hline
\multirow{3}{*}{Soil Moisture}
  & 10\% & 0.01 & 0.00 & 0.00 & 0.00 & 0.00 & 0.00 & 0.08 & 0.01 & \textbf{0.80} \\
  & 20\% & 0.01 & 0.00 & 0.00 & 0.00 & 0.01 & 0.01 & 0.01 & 0.01 & \textbf{0.80} \\
  & 30\% & 0.01 & 0.00 & 0.00 & 0.01 & 0.00 & 0.00 & 0.01 & 0.00 & \textbf{0.80} \\ \hline
\multirow{3}{*}{Tax}
  & 10\% & 0.13 & 0.01 & 0.02 & 0.00 & 0.00 & 0.00 & 0.01 & 0.00& \textbf{0.29} \\
  & 20\% & 0.25 & 0.02 & 0.11 & 0.00 & 0.00 & 0.00 & 0.01 & 0.00 & \textbf{0.34} \\
  & 30\% & 0.24 & 0.01 & 0.06 & 0.00 & 0.00 & 0.00 & 0.00 & 0.00 & \textbf{0.32} \\ \hline
\end{tabular}}
\caption{F$_1$ scores for error detection on tabular datasets corrupted with unified noise levels of $10\%$, $20\%$, and $30\%$, using the REIN error-injection strategy.}
\label{tab:unified_noise_10_20_30}
\end{table*}

\subsection{Ablation: Role of the Inlier Module}
We re-ran the pipeline (in Section \ref{sec:benchmark-clean-data-with-unified-noise}) with the statistical inlier filter turned off, keeping all downstream components unchanged. As shown in Table \ref{tab:inlier_ablation}, F\textsubscript{1} scores declined significantly across all datasets (having 30\% noise), emphasizing the filter’s critical role in providing clean context to the LLM to reduce false-positive rule generation.

\subsection{Cross-Domain Robustness Evaluation}


To gauge robustness across domains, we inject synthetic errors (using the \textbf{five-component error-injection} strategy in Section \ref{sec:benchmark-clean-data-with-unified-noise}) to four datasets that cover \textit{manufacturing} (3D Printer \cite{3DPrinter.dataset}), \textit{environmental monitoring} (Water-Quality \cite{Water.dataset}), \textit{scholarly communication} (Citation \cite{Movies.dataset}), and \textit{energy consumption} (Power \cite{Power.dataset}) with $30\%$ noise. We run the tests in two modes—one using domain-specific few-shot examples and one without—and report the resulting F\textsubscript{1} scores side-by-side in Table \ref{tab:domain_benchmark}, highlighting the merit of domain-specific examples.

\begin{table}[ht]
\centering
\resizebox{0.7\columnwidth}{!}{%
\begin{tabular}{|l|p{2cm}|p{2.5cm}|p{2.5cm}|}
\hline
Dataset & Noise \% & With Inlier Detection & Without Inlier Detection \\ \hline\hline
\multirow{3}{*}{Adult}
  & 10\% & \textbf{0.94} & 0.72 \\
  & 20\% & \textbf{0.80} & 0.77 \\
  & 30\% & \textbf{0.63} & 0.48 \\ \hline
\multirow{3}{*}{Beers}
& 10\% & \textbf{0.72} & 0.68 \\
  & 20\% & \textbf{0.82} & 0.68 \\
  & 30\% & \textbf{0.85} & 0.43 \\ \hline
\multirow{3}{*}{Bikes}
& 10\% & \textbf{0.82} & 0.81 \\
  & 20\% & \textbf{0.83} & 0.81 \\
  & 30\% & \textbf{0.82} & 0.81 \\ \hline
\multirow{3}{*}{Breast Cancer}
& 10\% & \textbf{0.90} & 0.85 \\
  & 20\% & \textbf{0.74} & 0.72 \\
  & 30\% & \textbf{0.89} & 0.72 \\ \hline
\multirow{3}{*}{HAR}
& 10\% & 0.42 & \textbf{0.51} \\
  & 20\% & \textbf{0.54} & 0.52 \\
  & 30\% & \textbf{0.55} & 0.53 \\ \hline
\multirow{3}{*}{Movies}
& 10\% & \textbf{0.71} & 0.69 \\
  & 20\% & \textbf{0.71} & 0.69 \\
  & 30\% & \textbf{0.68} & 0.66 \\ \hline
\multirow{3}{*}{Nasa}
& 10\% & \textbf{0.89} & 0.73 \\
  & 20\% & \textbf{0.85} & 0.71 \\
  & 30\% & \textbf{0.91} & 0.32 \\ \hline
\multirow{3}{*}{Rayyan}
& 10\% & \textbf{0.94} & \textbf{0.94} \\
  & 20\% & \textbf{0.97} & 0.94 \\
  & 30\% & \textbf{0.96} & 0.74 \\ \hline
\multirow{3}{*}{Soil Moisture}
& 10\% & \textbf{0.80} & 0.75 \\
  & 20\% & \textbf{0.80} & 0.55 \\
  & 30\% & \textbf{0.80} & 0.04 \\ \hline
\multirow{3}{*}{Tax}
& 10\% & \textbf{0.29} & 0.25 \\
  & 20\% & \textbf{0.34} & 0.23 \\
& 30 \% & \textbf{0.32} & 0.31 \\ \hline
\end{tabular}}
\caption{Ablation of F\textsubscript{1} scores across datasets ($30\%$ noise) with vs. without statistical inlier filter.}
\label{tab:inlier_ablation}
\end{table}


\subsection{Evaluation on Standard DQ Benchmarks}

We benchmark our framework on three data corruption suites, using the corresponding noisy datasets provided by each source. (i) We begin with the REIN benchmark \cite{REIN.Benchmark.Datasets}, which applies the \textbf{five-component error-injection} on six domain-diverse tables. Our results, compared with baselines including ED2, FAHES \cite{kdd18.FAHES}, KATARA, outlier detectors (IQR, IF, and SD), Max-Entropy, and Min-K, are presented in Table \ref{tab:rein_benchmark} (matching our reported results, \citet{SAGED.edbt24} also observe that the outlier detectors are not efficient on datasets like Beers, Flights, Hospital, and Nasa). (ii) We then evaluate on the six datasets from the ED2 benchmark \cite{ED2.Benchmark.Datasets}, using the same set of baselines; results appear in Table \ref{tab:ed2_raha_side_by_side}. (iii) Finally, we test on the five datasets from RAHA \cite{Raha.Benchmark.Datasets}, with comparative performance reported in Table \ref{tab:ed2_raha_side_by_side}. Our results for the datasets Beers, Flights, and Hospital differ on ED2 and Raha due to the difference in their noise.

\begin{table}[ht]
\centering
\resizebox{1.0\columnwidth}{!}{%
\begin{tabular}{|l|c|c|c|c|c|c|c|c|c|}
\hline
Dataset & ED2 & FAHES & KATARA & IQR & IF & SD & Max Entropy & Min-K & Ours \\
\hline
\hline
Adult            & 0.57 & 0.00 & 0.02 & 0.00 & 0.00 & 0.00 & 0.57 & 0.00 & \textbf{0.59}           \\
\hline
Beers           & 0.99 & 0.59 & 0.03 & 0.00 & 0.00 & 0.00 & 0.91 & 0.69 & \textbf{1.00} \\
\hline
Bikes             & 0.65 & 0.14 & 0.30 & 0.27 & 0.14 & 0.22 & 0.27 & 0.31 & \textbf{0.77}          \\
\hline
Breast Cancer   & 0.49 & 0.09 & 0.09 & 0.00 & 0.00 & 0.06 & 0.48 & 0.28 & \textbf{0.89}            \\
\hline
Flights          & 0.86 & 0.03 & 0.11 & 0.00 & 0.00 & 0.00 & 0.84 & 0.65 & \textbf{0.89}  \\
\hline
HAR               & 0.48 & 0.00 & 0.05 & 0.00 & 0.11 & 0.00 & 0.47 & 0.41 & \textbf{0.60}           \\
\hline
Hospital          & \textbf{0.99} & 0.01 & 0.08 & 0.00 & 0.00 & 0.00 & 0.74 & 0.5 & 0.86  \\
\hline
Mercedes          & 0.32 & 0.00 & 0.00 & 0.00 & 0.01 & 0.01 & 0.21 & 0.00 &  \textbf{0.73}           \\
\hline
Nasa             & 0.76 & 0.05 & 0.13 & 0.00 & 0.00 & 0.00 & 0.32 & 0.22 & \textbf{0.96}          \\
\hline
Soil Moisture     & 0.05 & 0.00 & 0.00 & 0.00 & 0.04 & 0.02 & 0.03 & 0.03 &  \textbf{0.59}        \\
\hline
\end{tabular}}
\caption{REIN Benchmark Evaluation: Error Detection Accuracy (F\textsubscript{1}) of Our Approach vs. Baselines}
\label{tab:rein_benchmark}
\end{table}



\begin{table}[ht]
\centering\small
\begin{tabular}{@{}c c@{}}   
\scalebox{0.9}{   
\begin{tabular}{|l|c|c|}
\hline
\textbf{Dataset} & \textbf{ED2} & \textbf{Ours} \\ \hline\hline
Beers      & 0.98 & \textbf{1.00} \\ \hline
Flights    & 0.86 & \textbf{0.88} \\ \hline
Hospital   & \textbf{1.00} & 0.82 \\ \hline
Restaurant & \textbf{0.76} & 0.61 \\ \hline
Soccer     & \textbf{0.81} & \textbf{0.81} \\ \hline
\end{tabular}}
&
\scalebox{0.9}{
\begin{tabular}{|l|c|c|}
\hline
\textbf{Dataset} & \textbf{RAHA} & \textbf{Ours} \\ \hline\hline
Beers    & \textbf{0.99} & 0.95 \\ \hline
Flights  & 0.82 & \textbf{0.89} \\ \hline
Hospital & 0.75 & \textbf{0.86} \\ \hline
Movies   & 0.84 & \textbf{0.85} \\ \hline
Rayyan   & \textbf{0.78} & 0.75 \\ \hline
\end{tabular}}
\end{tabular}
\caption{Left: ED2 benchmark; Right: RAHA benchmark (F\textsubscript{1} scores).}
\label{tab:ed2_raha_side_by_side}
\end{table}

\textbf{Observations:}
Across ten data sets and three noise levels, our method records the top F\textsubscript{1} in 26 of 30 settings, markedly surpassing the next best detector on challenging tables such as Bikes, Breast Cancer, and Soil-Moisture. With domain-tailored few-shots, the Power-Consumption table's F\textsubscript{1} rises from 0.61 to 0.71, showing a clear gain over the generic prompt. On the REIN benchmark our system posts the top score on 9 of 10 data sets—often by wide margins on Bikes, Breast Cancer and Soil Moisture—while conceding only Hospital to ED2.

    \begin{table}[ht]
    \centering
    \small
    \resizebox{0.9\columnwidth}{!}{%
    \begin{tabular}{|l|p{2cm}|p{2.5cm}|}
    \hline
    \multirow{2}{*}{Dataset} & \multicolumn{2}{c|}{Few-shot examples} \\ \cline{2-3}
                             & With domain-specific & Without domain-specific \\ \hline\hline
    3D Printer        & \textbf{0.68} & 0.66 \\ \hline
    Citation          & \textbf{0.80} & 0.78 \\ \hline
    Power-Consumption & \textbf{0.71} & 0.61 \\ \hline
    Water-Quality     & \textbf{0.21} & 0.17 \\ \hline
    \end{tabular}}
    \caption{Error detection F\textsubscript{1} scores with vs. without domain-specific few-shot examples.}
    \label{tab:domain_benchmark}
    \end{table}

\section{Conclusion}

We introduced a fully automated, three-stage framework that integrates large-scale statistical inlier detection, LLM-based semantically valid rule generation, and code synthesis to produce executable data-quality validators for tabular datasets. Each quality rule is encapsulated as a structured, human-readable rule card, promoting transparency and expert oversight. Conflicts among generated DQ rules are resolved agnostically to rule types and consolidated using a rubric for consistency. Our approach scales to large tabular datasets by eliminating the need for cell-level LLM inference calls. Evaluated on both REIN synthetic stress benchmarks and standard datasets from ED2 and RAHA, our approach consistently outperforms existing detectors in accuracy. By incorporating external domain knowledge and domain-specific few-shot examples in the prompts, our approach allows reliable and domain-agnostic data quality assurance.
\newpage
\bibliography{acl_latex}
\newpage
\appendix
\section{Appendix}
\label{sec:appendix}
\subsection{Confidence Analysis with and without few-shot examples}
\label{sec:confidence_analysis}

To assess the reliability of the rules generated by the large language model (LLM), we conducted a systematic confidence analysis across four datasets: Beers \cite{Beers.dataset}, SmartFactory \cite{Smartfactory.dataset}, NASA \cite{Nasa.dataset}, and Adult \cite{Adult.dataset} under two prompting conditions: (i) with few-shot exemplars and (ii) without few-shot exemplars. Our hypothesis was that providing few-shot guidance would not only improve the overall quality of generated rules but also lead to higher model confidence and broader schema coverage.

For each given prompt LLM produces multiple candidate rules per rule type. To make comparisons fair across datasets and rule types, we evaluated the \textbf{confidence} and \textbf{coverage} for each LLM response and then aggregated across datasets. 

\begin{itemize}
  \item \textbf{Per-rule Confidence:} We captured the log probabilities produces at token level in a the generated rule. The obtained log-probabilities are then used to compute Linear probability.
  
  \item \textbf{Coverage of Target Columns: } To measure the breadth of the generated rules, we defined coverage as the proportion of schema columns touched by at least one rule (where we consider the final set of rules post the guardrail filter processing). Given $X$ as the set of unique target columns referenced in generated rules and $Y$ as the set of columns in the dataset schema, coverage was computed as:
  \begin{equation}
      \text{Coverage} = \frac{|X \cap Y|}{|Y|}
  \end{equation}
      
  \item \textbf{Dataset-level Aggregation: } Because the number of rules varied across conditions and datasets, we adopted a macro-averaging strategy: coverage and confidence were computed per dataset and per rule type, then averaged across datasets. This normalization ensures that datasets with larger schema or more generated rules do not dominate the analysis.
  
\end{itemize}

The results of confidence and coverage of different rule types are shown in Table \ref{tab:confidence_coverage}. Confidence values are averaged linear probabilities expressed on a scale from 0 to 100, where higher values indicate greater model certainty. Coverage values range from 0 to 1 and represent the fraction of schema columns targeted by at least one generated rule.

\begin{table*}[ht]
\centering
\resizebox{0.85\textwidth}{!}{%
\begin{tabular}{|l||c|c|c|c|}
\hline
\textbf{Rule Type} & 
\textbf{Conf. (Linear, With FS)} & 
\textbf{Conf. (Linear, Without FS)} & 
\textbf{Coverage (With FS)} & 
\textbf{Coverage (Without FS)} \\
\hline
CROSS\_COLUMN\_VALIDATION     & 96.17 & 91.67 & 0.494 & 0.480 \\ \hline
DATA\_TYPE\_VALIDATION        & 98.23 & 93.73 & 0.848 & 0.875 \\ \hline
DEPENDENCY\_CONSTRAINTS       & 97.03 & 91.14 & 0.665 & 0.565 \\ \hline
FORMAT\_COMPLIANCE            & 96.40 & 92.85 & 0.678 & 0.278 \\ \hline
RANGE\_CONSTRAINTS            & 96.13 & 81.95 & 0.490 & 0.324 \\ \hline
TEMPORAL\_CONSISTENCY\_CHECKS & 97.20 & 86.69 & 0.510 & 0.414 \\ \hline \hline
\textbf{Average}              & 96.86 & 89.67 & 0.614 & 0.489 \\ \hline
\end{tabular}}
\caption{Average rule confidence (linear probabilities) and target-column coverage across Beers, SmartFactory, NASA, and Adult datasets. 
Results are reported per rule type under two prompting conditions: \emph{with few-shot examples} and \emph{without few-shot examples}. 
Coverage is normalized by schema size.}
\label{tab:confidence_coverage}
\end{table*}


\subsection{Background of DQ Rules}
\label{sec:dq_rule_background}
Table \ref{tab:dirty-data-sample} presents an example dataset illustrating different types of data quality issues discussed in our paper.
\begin{itemize}[leftmargin=*]
    \item {\bf Reference Table Validation}: Each ZIP Code in the Warehouse column should be a valid entry of the reference table with postal codes.
    \item {\bf Missing Value Identification}: The ProductName value is missing for the record with ProductID {\it P2002}.
    \item {\bf Pattern Matching}: The LastRestock date for the record with ProductID {\em P2004} does not follow the {\em YYYY-MM-DD} format.
    \item {\bf Value Set Constraint}: Category column cannot allow values outside of set $\{\text{Electronics}, \text{Accessories}\}$.
    \item {\bf Range Constraints}: Stock TurnoverRate must consist of values in the range [0, 100]. Quantity should be a positive integer. 
    \item {\bf Uniqueness Constraint}: The ProductID value {\it P2001} should not repeat.
    \item {\bf Format Compliance}: The ProductID should be of the form {\it P} $([0-9])\{4\}$ , e.g: P1011, P3002, etc.
    \item {\bf Data Type Validation}: The `LastRestock' column must be a date type. 
\end{itemize}

\begin{table*}[ht!]
\centering
\small
\begin{tabular}{|c|c|c|c|c|c|c|c|}
\hline
ProductID & ProductName & Price & Quantity & Category & LastRestock & Warehouse & TurnoverRate \\
\hline
\hline
P2001 & Laptop Pro 15 & 1299.99 & 50 & Electronics & 2024-12-15 & \textcolor{magenta}{948102} & 45.0 \\
\hline
P2002 &  & 89.99 & 200 & Accessories & \textcolor{blue}{2025-30-10} & 30301 & 82.5 \\
\hline
\textcolor{purple}{P2001} & Phone Charger & 99999.50 & 300 & \textcolor{violet}{Bike} & 2025-06-15 & 60601 & 68.5 \\
\hline
P2004 & Bluetooth Speaker & 45.50 & \textcolor{olive}{-5} & Electronics & \textcolor{orange}{03-14025} & 60601 & 56.5 \\
\hline
\textcolor{red}{PPX2001} & Laptop Pro 15 & 1299.99 & 50 & Electronics & 2024-12-15 & 30301 & 75.0 \\
\hline
P2005 & Gaming Mouse & 59.99 & 100 & Electronics & 2025-05-20 & 60601 & \textcolor{brown}{9059.5} \\
\hline
\end{tabular}
\caption{A sample Product Inventory dataset. Error cells are colored.}
\label{tab:dirty-data-sample}
\end{table*}


\begin{table*}[ht!]
\centering
\begin{tabular}{||p{2.4cm}||p{4.4cm}|p{8.5cm}||}
\hline
\textbf{DQ Dimension}        & \textbf{DQ Rule Type}                 & \textbf{Sample DQ Rules (Rule Cards)}  \\
\hline
\multirow{2}{*}{Conformity} & \multirow{2}{*}{Pattern Matching}     & $\forall x \in \texttt{article\_pagination},\quad x \models \texttt{[0-9]+\text{-}[0-9]+}$                                                                            \\ \cline{3-3} 
                             &                                       & $\forall x \in \texttt{article\_language},\quad x \models \texttt{[a-z]\{3\}}$  \\ 
\hline
\multirow{9}{*}{Validity}                     & \multirow{2}{*}{Business rule validation} & $Discount \leq 0.2 \times Order\_Value$                                              \\ \cline{3-3} 
                                                &         & $ \newline \frac{Discount\_Amount}{Original\_price} \leq 0.5 \newline $ \\
                                                \cline{2-3}
                                                & \multirow{1}{*}{Computation Consistency} & $Order\_Total = \sum_{i=1}^{n}{Item\_Price_{i} \times Quantity_{i}}$    
                                                \\
                                                \cline{2-3} 
                                                & \multirow{2}{*}{Dependency Constraints}                                              & $order\_status == \text{'Returned'} \Rightarrow return\_initiation\_date \neq \text{NULL}$            \\
                                                \cline{2-3}
                                                & \multirow{2}{*}{Range Constraints} & $0\% \leq discount \leq 90\%$                                              \\ \cline{3-3} 
                                                &                                          & $6 \leq warranty\_months \leq 60 $ \\ \cline{2-3}
                                                & \multirow{2}{*}{Temporal Consistency} & $order\_date \leq estimated\_delivery\_date$                                              \\ \cline{3-3} 
                                                &                                          & $actual\_delivery\_date \leq return\_date$
                                                \\
                                                \hline
\end{tabular}
\caption{Additional examples of DQ Rules categorized by  DQ Dimensions and DQ Rule Types (Movies dataset).}
\label{tab:add_dq_rules}
\end{table*}

\subsection{Examples of Conflicting Rules}
\label{sec:examples-of-conflicting-rules}
\begin{figure}[h]
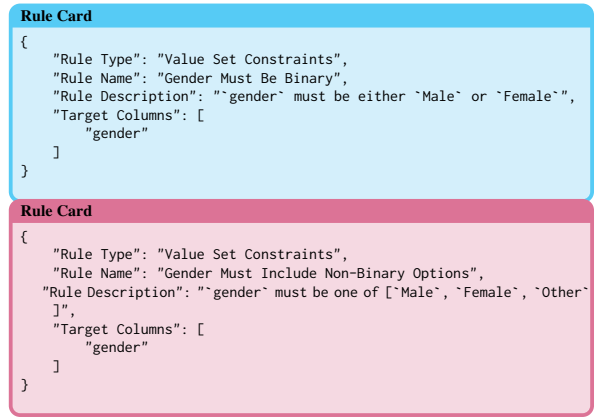

    \tiny
    \setlength{\topsep}{0pt}
    \setlength{\partopsep}{0pt}
    \setlength{\parskip}{0pt}
    \centering
    \begin{tcolorbox}[
        colback=cyan!15!white,
        colframe=cyan!55!white,
        title=\textcolor{black}{\textbf{Rule Card}},
        boxsep=1pt,
        left=2pt,
        right=2pt,
        top=2pt,
        bottom=1pt,
        width=\linewidth
    ]
    \begin{verbatim}
{
    "Rule Type": "Value Set Constraints",
    "Rule Name": "Gender Must Be Binary",
    "Rule Description": "`gender` must be either `Male` or `Female`",
    "Target Columns": [
        "gender"
    ]
}
    \end{verbatim}
    \end{tcolorbox}
    \vspace{-3mm}
    \begin{tcolorbox}[
        colback=purple!15!white,
        colframe=purple!55!white,
        title=\textcolor{black}{\textbf{Rule Card}},
        boxsep=1pt,
        left=2pt,
        right=2pt,
        top=2pt,
        bottom=1pt,
        width=\linewidth
    ]
    \begin{verbatim}
{
    "Rule Type": "Value Set Constraints",
    "Rule Name": "Gender Must Include Non-Binary Options",
    "Rule Description": "`gender` must be one of [`Male`, `Female`, `Other`
    ]",
    "Target Columns": [
        "gender"
    ]
}
    \end{verbatim}
    \end{tcolorbox}
    \caption{Conflicting rules due to inconsistencies in the sampled data frames of the Hospital dataset \cite{ED2.Benchmark.Datasets}}
    \label{fig:conflict_due_to_dataset_sampling}
\end{figure}

\begin{figure}[h]
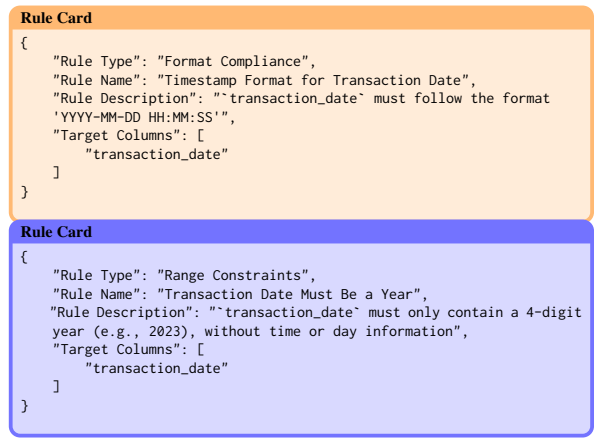

    \tiny
    \setlength{\topsep}{0pt}
    \setlength{\partopsep}{0pt}
    \setlength{\parskip}{0pt}
    \centering
    \begin{tcolorbox}[
        colback=orange!15!white,
        colframe=orange!55!white,
        title=\textcolor{black}{\textbf{Rule Card}},
        boxsep=1pt,
        left=2pt,
        right=2pt,
        top=2pt,
        bottom=1pt,
        width=\linewidth
    ]
    \begin{verbatim}
{
    "Rule Type": "Format Compliance",
    "Rule Name": "Timestamp Format for Transaction Date",
    "Rule Description": "`transaction_date` must follow the format 
    'YYYY-MM-DD HH:MM:SS'",
    "Target Columns": [
        "transaction_date"
    ]
}
    \end{verbatim}
    \end{tcolorbox}
    \vspace{-3mm}
    \begin{tcolorbox}[
        colback=blue!15!white,
        colframe=blue!55!white,
        title=\textcolor{black}{\textbf{Rule Card}},
        boxsep=1pt,
        left=2pt,
        right=2pt,
        top=2pt,
        bottom=1pt,
        width=\linewidth
    ]
    \begin{verbatim}
{
    "Rule Type": "Range Constraints",
    "Rule Name": "Transaction Date Must Be a Year",
    "Rule Description": "`transaction_date` must only contain a 4-digit 
    year (e.g., 2023), without time or day information",
    "Target Columns": [
        "transaction_date"
    ]
}
    \end{verbatim}
    \end{tcolorbox}
    \caption{Conflicting rules due to language model hallucinations of the Hospital dataset \cite{ED2.Benchmark.Datasets}}
    \label{fig:conflict_due_to_model_hallucination}
\end{figure}

Conflicts may arise both within and across rule types among the generated DQ rules.
Figure \ref{fig:conflict_due_to_dataset_sampling} depicts contradictory rules of the same rule type arising from inconsistencies in the sample data frames used across iterations (e.g., one sample might include only `Male' and `Female' in a gender column, while another might also include `Other') during the generation of the rule cards.

Figure \ref{fig:conflict_due_to_model_hallucination} depicts contradictory rules across rule types arising from language model hallucinations (e.g., conflicting rules such as one requiring transaction\_date to follow the `YYYY-MM-DD' format, and another limiting it to just a $4$-digit year).

We could resolve conflicts among DQ rules either by incorporating user preferences on the priorities of DQ rule types, or by using the reasoning capabilities of LLMs to retain only the semantically relevant DQ rules, or even by dropping DQ rules one by one that are in conflict.

\begin{figure}[h]
    \tiny
    \setlength{\topsep}{0pt}
    \setlength{\partopsep}{0pt}
    \setlength{\parskip}{0pt}
    \centering
    \begin{tcolorbox}[
        colback=cyan!15!white,
        colframe=cyan!55!white,
        title=\textcolor{black}{\textbf{Table Schema:}},
        boxsep=1pt,
        left=2pt,
        right=2pt,
        top=2pt,
        bottom=1pt,
        width=\linewidth
    ]
    \begin{verbatim}
CREATE TABLE  ( instant INTEGER -- Index of the data in the 
      dataset. Eg. 1, 11592, 11578, 11579, 11580  
        dteday DOUBLE -- time interval of the given data. Eg. 
        359399.01666666666, 368207.06666666665, 368807.06666666665, 
        368831.06666666665, 368855.06666666665  
        season INTEGER -- seasonal attribute. Belongs to the set: [1, 2, 
        3, 4]  
        yr INTEGER -- year attribute of the data. Belongs to the set: 
        [0, 1]  
        mnth INTEGER -- month of the data. Eg. 5, 7, 12, 8, 3  
        hr INTEGER -- hour of the data. Eg. 17, 16, 13, 15, 14  
        holiday INTEGER -- whether day was holiday or not. Belongs to 
        the set: [0, 1]  
        weekday INTEGER -- which day of the week. Belongs to the set: 
        [6, 0, 1, 2, 3, 4, 5]  
        workingday INTEGER -- whether it is working day or not. Belongs 
        to the set: [0, 1]  
        weathersit INTEGER -- attribute describing weather conditions. 
        Belongs to the set: [1, 2, 3, 4]  
        temp DOUBLE -- temperature of the day. Eg. 0.62, 0.66, 0.64, 
        0.7, 0.6  
        atemp DOUBLE -- temperature in the morning of the day. Eg. 
        0.6212, 0.5152, 0.4091, 0.3333, 0.6667  
        hum DOUBLE -- humidity of the day. Eg. 0.88, 0.83, 0.94, 0.87, 
        0.7  
        windspeed DOUBLE -- speed of the wind. Eg. 0.0, 0.1343, 0.1642, 
        0.194, 0.1045  
        casual INTEGER -- type of customer. Eg. 0, 1, 2, 3, 4          
        registered INTEGER -- users that have registered through 
        application. Eg. 4, 3, 5, 6, 2  
        cnt INTEGER -- count of something that is recorded. Eg. 5, 6, 4, 
        3, 2  
  )
    \end{verbatim}    
    \end{tcolorbox}  
    \caption{An example schema generated by the system on Bike dataset \cite{Bikes.dataset}.}
    \label{fig:schema_example}
\end{figure}

\begin{figure}[h]
    \tiny
    \setlength{\topsep}{0pt}
    \setlength{\partopsep}{0pt}
    \setlength{\parskip}{0pt}
    \centering
    \begin{tcolorbox}[
        colback=cyan!15!white,
        colframe=cyan!55!white,
        title=\textcolor{black}{\textbf{Column Profile:}},
        boxsep=1pt,
        left=2pt,
        right=2pt,
        top=2pt,
        bottom=1pt,
        width=\linewidth
    ]
    \begin{verbatim}
 {
  "Name": "weekday",
  "Expected Type": "int",
  "Unique Values": [0, 1, 2, 3, 4, 5, 6],
  "distinct_count": 7,
  "Min Value": 0.0,
  "Max Value": 6.0,
  "Duplicates %": 99.96
}
    \end{verbatim}    
    \end{tcolorbox}  
    \caption{An example column profile for weekday generated by the system on Bike dataset \cite{Bikes.dataset}.}
    \label{fig:col_summary}
\end{figure}

\subsection{Prompts used in our End-to-End pipeline system}
\label{sec:Appendix_end_to_end_pipeline}

\begin{figure}[h]
    \tiny
    \setlength{\topsep}{0pt}
    \setlength{\partopsep}{0pt}
    \setlength{\parskip}{0pt}
    \centering
    \begin{tcolorbox}[
        colback=cyan!15!white,
        colframe=cyan!55!white,
        title=\textcolor{black}{\textbf{Task:}},
        boxsep=1pt,
        left=2pt,
        right=2pt,
        top=2pt,
        bottom=1pt,
        width=\linewidth
    ]
    \textit{
    Generate \textbf{Format Compliance} for the schema below. A rule must flag any value that does not match the expected data type of its column  and may add length, format, or categorical constraints when appropriate.  Avoid vague comparisons (e.g.\ “higher” without a reference point) and do not impose string rules on numeric columns.
    } \\
    \textit{ \\
    What is Format Compliance?  \\
    - Some columns must follow a specific format (e.g., phone numbers, 
    email addresses, date formats, or identification numbers). \\
    - If a value does not match the expected format, it should be 
    flagged as invalid.  \\
    - Format compliance rules help maintain consistency and enable 
    seamless data processing.  \\
    - Do Not enforces a strict format on a naturally diverse column 
    (e.g., names, addresses, product descriptions).  \\
    - Do Not impose arbitrary constraints without a clear logical basis 
    (e.g., restricting email domains, forcing specific city names).  \\
    - Do Not rule unnecessarily limits valid values where variation is 
    expected (e.g., requiring specific phone numbers).  
    }
    \end{tcolorbox}
    \vspace{-3mm}
    \begin{tcolorbox}[
        colback=blue!15!white,
        colframe=blue!45!white,
        title= \textcolor{black}{\textbf{Example Schema:}},
        boxsep=1pt,
        left=2pt,
        right=2pt,
        top=2pt,
        bottom=0pt,
        width=\linewidth
    ]
        \begin{verbatim}
```sql
CREATE TABLE employee_records (
  employee_id VARCHAR(10)  -- "EMP12345"
  email        VARCHAR(255) -- john.doe@example.com
  phone_number VARCHAR(15)  -- +1-202-555-0173
  date_of_birth DATE        -- 1985-06-15
  postal_code  VARCHAR(10)  -- 10001 / SW1A 1AA
  ssn          VARCHAR(11)  -- 123-45-6789
)
```
    \end{verbatim}
    \end{tcolorbox}
    \vspace{-4mm}
     \begin{tcolorbox}[
        colback=purple!5!white,
        colframe=purple!25!white,
        title=\textcolor{black}{\textbf{Example Format Compliance Rules:}},
        boxsep=1pt,
        left=2pt,
        right=2pt,
        top=2pt,
        bottom=0pt,
        width=\linewidth
    ]
        \begin{verbatim}
```json
[
  {"Rule Name":"Employee ID Follows 'EMP' + Digits",
   "Rule Description":"`employee_id` must begin with 'EMP' and
           be followed by digits (e.g., EMP12345).",
   "Target Columns":["employee_id"]},
  {"Rule Name":"Email Must Follow Standard Format",
   "Rule Description":"`email` must contain one '@' and a valid
           domain (e.g., john.doe@example.com).",
   "Target Columns":["email"]}, ..
```
    \end{verbatim}
    \end{tcolorbox}
    \vspace{-4mm}
     \begin{tcolorbox}[
        colback=orange!5!white,
        colframe=orange!25!white,
        title=\textcolor{black}{\textbf{Domain Specific Few-Shot Examples:}},
        boxsep=1pt,
        left=2pt,
        right=2pt,
        top=2pt,
        bottom=0pt,
        width=\linewidth
    ]
    \begin{verbatim}
```json
[
  {
    "Rule Name": "ICD-10 Code Must Follow Standard Pattern",
    "Rule Description": "The `icd10_code` column must match the ICD-10
    format: one uppercase letter, two digits, optionally a period and
    one to four alphanumeric characters (e.g., 'E11.9', 'M54.50').
    Any value outside this pattern should be flagged as invalid.",
    "Target Columns": ["icd10_code"]
  }...
    \end{verbatim}
    \end{tcolorbox}
    
    \vspace{-4mm}
     \begin{tcolorbox}[
        colback=purple!5!white,
        colframe=purple!25!white,
        title=\textcolor{black}{\textbf{Rules From Previous Iterations For Test Schema:}},
        boxsep=1pt,
        left=2pt,
        right=2pt,
        top=2pt,
        bottom=0pt,
        width=\linewidth
    ]
    \begin{verbatim}
    ```json
[
{
    "Rule Name": "State Must Be Two-Letter Lowercase Code",
    "Rule Description": "The `State` column must contain a valid two-letter 
    U.S. state abbreviation in lowercase (e.g., 'al', 'ak').",
    "Target Columns": ["State"]
  },
  {
    "Rule Name": "Phone Number Must Be 10 Digits",
    "Rule Description": "The `PhoneNumber` column must contain exactly 
    ten numeric digits with no separators (e.g., '2053258100').",
    "Target Columns": ["PhoneNumber"]
  }...
    \end{verbatim}
    \end{tcolorbox}
    
    \vspace{-4mm}
     \begin{tcolorbox}[
        colback=olive!5!white,
        colframe=olive!25!white,
        title=\texttt{Instructions:},
        boxsep=1pt,
        left=2pt,
        right=2pt,
        top=2pt,
        bottom=0pt,
        width=\linewidth
    ]
        \texttt{\\
        Instructions:  \\
        1. Use the schema provided to generate format compliance 
        rules.  \\
        2. Follow the example format to define how values should be 
        structured.  \\
        3. Ensure that every column with a predefined format has a 
        corresponding rule.  \\
        4. Use clear descriptions so that the format requirements are 
        well understood.  \\
        5. MAKE SURE TO GENERATE PROPER JSON FORMAT. \\
        Write atleast 15 rules.
        Do not write anything other than rules.
         }
    \end{tcolorbox}
    \vspace{-3mm}
     \begin{tcolorbox}[
        colback=orange!15!white,
        colframe=orange!55!white,
        title=\texttt{Test Input:},
        boxsep=1pt,
        left=2pt,
        right=2pt,
        top=2pt,
        bottom=0pt,
        width=\linewidth
    ]
    \texttt{Test Schema: \\
    ```sql \\
    ... \\
    ``` \\
    Start generating 15 rules in the specified JSON format.
    }
    \end{tcolorbox}
    \caption{Prompt Template to generate Rule Cards for Rule Type `Format Compliance' for Beers dataset \cite{Beers.dataset}}
    \label{fig:rule_prompt_format_compliance}
\end{figure}

\begin{figure}[h]
    \tiny
    \setlength{\topsep}{0pt}
    \setlength{\partopsep}{0pt}
    \setlength{\parskip}{0pt}
    \centering
    \begin{tcolorbox}[
        colback=blue!5!white,
        colframe=blue!35!white,
        title=\textcolor{black}{\textbf{Rule Card 1:}},
        boxsep=1pt,
        left=2pt,
        right=2pt,
        top=2pt,
        bottom=1pt,
        width=\linewidth
    ]
    \begin{verbatim}
{
    "Rule Name": "ABV Must Follow 0.XXX Decimal Format",
    "Rule Description": "The `abv` column must be a decimal between 0 and 
    1 with a leading zero and up to three decimal places (e.g., 0.050, 
    0.090).",
    "Target Columns": ["abv"]
  }
    \end{verbatim}
    \end{tcolorbox}
     \vspace{-3mm}
    \begin{tcolorbox}[
        colback=orange!5!white,
        colframe=orange!35!white,
        title=\textcolor{black}{\textbf{Rule Card 2:}},
        boxsep=1pt,
        left=2pt,
        right=2pt,
        top=2pt,
        bottom=1pt,
        width=\linewidth
    ]
    \begin{verbatim}
{
    "Rule Name": "State Must Be Two-Letter Uppercase Code",
    "Rule Description": "The `state` column must contain a valid two-
    letter U.S. state abbreviation in uppercase (e.g., \"OR\", \"IN\").",
    "Target Columns": ["state"]
  }
    \end{verbatim}
    \end{tcolorbox}
 \vspace{-3mm}
    \begin{tcolorbox}[
        colback=green!5!white,
        colframe=green!35!white,
        title=\textcolor{black}{\textbf{Rule Card 3:}},
        boxsep=1pt,
        left=2pt,
        right=2pt,
        top=2pt,
        bottom=1pt,
        width=\linewidth
    ]
    \begin{verbatim}
{
    "Rule Name": "Beer Name Should Contain Letters, Numbers, or Spaces",
    "Rule Description": "The `beer-name` column may include letters, 
    numbers, apostrophes, or spaces but no other special characters.",
    "Target Columns": ["beer-name"]
  }
    \end{verbatim}
    \end{tcolorbox}
    \caption{Generated rule cards for Rule Type `Format Compliance' for Beers dataset \cite{Beers.dataset}}
    \label{fig:rule_cards_beers_dataset}
\end{figure}

\begin{figure*}[ht]
  \tiny
  \centering
  \begin{tcolorbox}[colback=cyan!15!white,colframe=cyan!55!white,
                    title=\textcolor{black}{\textbf{Task:}},
                    boxsep=1pt,left=2pt,right=2pt,top=2pt,bottom=1pt,
                    width=\linewidth]
\texttt{TASK \\
For the given COMPARE\_TABLE, COLUMN\_PROFILE, and RULE:\\
1. Decide the rule’s usefulness and assign a rule **Status** from the table.\\
2. Explain briefly ($\leq$ 30 words).\\
3. Always supply `Additional Information` (see rules below).}

\texttt{\\| Status label | When to use |\\
|-------|-------------|\\
| **correct**               | Rule fits the profile and adds value. |\\
| **incomplete**            | Rule is conceptually right but missing details.| \\
| **incorrect\_fixable**     | Rule conflicts with data but can be repaired. |\\
| **incorrect\_not\_fixable** | Rule conflicts with data and cannot be salvaged. |\\
| **irrelevant**            | Rule adds no value (e.g., metadata only). |\\
| **redundant**             | Rule duplicates another (rare in single-rule mode). |\\
| **unimplementable**       | Requires external data that is unavailable. |}


\end{tcolorbox}

\vspace{-3mm}
  \begin{tcolorbox}[colback=pink!15!white,colframe=pink!45!white,
                    title=\textcolor{black}{\textbf{Additional Information Requirements}},
                    boxsep=1pt,left=2pt,right=2pt,top=2pt,bottom=0pt,
                    width=\linewidth]
\texttt{Additional Information requirements\\
- If status is correct or incorrect\_fixable, provide the Specification (optional) and Pseudocode bullets:}
\begin{verbatim}
  "Additional Information": { "Specification": "<clear rule text>", "Pseudocode": ["condition 1 → flag", "condition 2 → flag"]}
\end{verbatim}
\texttt{Otherwise use the `DROP RULE – …' string in Specification.}\\
\texttt{\\Value-set constraints:\\
- Only propose an explicit list of allowed values if the column’s `distinct\_count' is $\leq$ 30 (or the `Unique Values' sample shows 30 or fewer items).\\
- Otherwise:\\
* mark the rule **incomplete** (needs different constraint), *or*\\
* propose a pattern / range / format check instead.\\
\\Pattern Matching constraints:\\
- Do \emph{not} apply regex constraints to purely numeric columns;\\
- Avoid over-fitting: include only character classes that appear in all clean examples and omit hard-coded constants unless they are present in every value.\\
\\Cross-column validation:\\
- Do not duplicate single-column checks already covered elsewhere; mark such cases redundant.\\
- Include only columns that co-occur in at least 90 \% of sample rows; otherwise label incomplete.}
\\


\end{tcolorbox}

\vspace{-3mm}
  \begin{tcolorbox}[colback=blue!15!white,colframe=blue!45!white,
                    title=\textcolor{black}{\textbf{Few-shot Example COLUMN\_PROFILE}},
                    boxsep=1pt,left=2pt,right=2pt,top=2pt,bottom=0pt,
                    width=\linewidth]
                    
\texttt{Few-shot Example 1 — Incomplete rule fixed}\\
\texttt{COLUMN\_PROFILE}
\begin{verbatim}
{
  "Name": "flight",
  "Expected Type": "str",
  "Pattern": "Unrecognized",
  "distinct_count": 100,
  "Some Unique Values": ['AA-3859-IAH-ORD', 'AA-1733-ORD-PHX', 'AA-1640-MIA-MCO', 'AA-518-MIA-JFK', 'AA-3756-ORD-SLC', 'AA-204-LAX-MCO', 
  'AA-3468-CVG-MIA', 'AA-484-DFW-MIA', 'AA-446-DFW-PHL', 'AA-466-IAH-MIA', 'AA-1886-BOS-MIA', 'AA-2957-DFW-CVG', 'AA-1664-MIA-ATL', 
  'AA-3979-CVG-ORD', 'AA-1279-DFW-PHX', 'AA-616-DFW-DTW', 'AA-4344-ORD-DTW', 'AA-2525-DFW-MIA', 'AA-404-MIA-MCO']
}
\end{verbatim}
\end{tcolorbox}

\vspace{-3mm}
  \begin{tcolorbox}[colback=orange!15!white,colframe=orange!45!white,
                    title=\textcolor{black}{\textbf{Few-shot Example COMPARE\_TABLE}},
                    boxsep=1pt,left=2pt,right=2pt,top=2pt,bottom=0pt,
                    width=\linewidth]
\texttt{COMPARE\_TABLE}
\texttt{\\ | Correct     | Noise      |\\
| --------------- | -------------- |\\
| AA-3823-LAX-DEN | AA3823-LAX-DEN |\\
| AA-2312-DFW-DTW | AA-2312DFW-DTW |\\
| AA-1165-JFK-MIA | AA-1165-JFKMIA |\\
| AA-431-MIA-SFO  | AA431-MIA-SFO  |\\
| AA-649-ORD-SNA  | AA-649-ORDSNA  |\\
| AA-3063-SLC-LAX | AA-3063SLC-LAX |}
\end{tcolorbox}

\vspace{-3mm}
  \begin{tcolorbox}[colback=purple!15!white,colframe=purple!45!white,
                    title=\textcolor{black}{\textbf{Few-shot Example}},
                    boxsep=1pt,left=2pt,right=2pt,top=2pt,bottom=0pt,
                    width=\linewidth]

\texttt{RULE}
\begin{verbatim}
{
  "Rule Name": "Flight Must Be a Valid String",
  "Rule Description": "The `flight` column must contain only valid string values.",
  "Target Columns": ["flight"]
}
\end{verbatim}
\texttt{EXPECTED JSON}
\begin{verbatim}
```json
{
  "Rule Name": "Flight Must Be a Valid String",
  "Status": "incorrect_fixable",
  "Reason": "Needs concrete airline-number-origin-dest pattern.",
  "Additional Information": {
    "Specification": "flight must match airline-number-origin-dest pattern.",
    "Pseudocode": [
      "if flight is null → flag",
      "if flight does not match ^[A-Z]\\{2\\}-\\d{1,4}-[A-Z]{3}-[A-Z]{3} → flag"
    ]
  }
}
```    
\end{verbatim}
\end{tcolorbox}

\vspace{-3mm}
\begin{tcolorbox}[colback=purple!5!white,colframe=purple!25!white,
title=\textcolor{black}{\textbf{Test}},
boxsep=1pt,left=2pt,right=2pt,top=2pt,bottom=0pt,
width=\linewidth]
\texttt{COMPARE\_TABLE ...} \\
\texttt{COLUMN\_PROFILE ...} \\
\texttt{RULE} 
\begin{verbatim}
{
  "Rule Type": "Format Compliance",
  "Rule Name": "Flight Must Follow Airline-FlightNumber-Origin-Destination Format",
  "Rule Description": "The `flight` column must follow the format: 
            Airline Code (2-3 letters) - Flight Number (1-4 digits) - Origin Airport Code (3 letters) - Destination Airport Code (3 letters). 
            Examples include 'AA-59-JFK-SFO', 'UA-664-ORD-PHL'. Any values not conforming to this pattern should be flagged as invalid.",
  "Target Columns": ["flight"]
}
\end{verbatim}
\end{tcolorbox}
  \caption{Prompt template to enrich Rule Cards  (agnostic to Rule Type).}
  \label{fig:enrich_rule_prompt_template}
\end{figure*}

\begin{figure}[h]
    \tiny
    \setlength{\topsep}{0pt}
    \setlength{\partopsep}{0pt}
    \setlength{\parskip}{0pt}
    \centering
    \begin{tcolorbox}[
        colback=blue!5!white,
        colframe=blue!35!white,
        title=\textcolor{black}{\textbf{Rule Card 1:}},
        boxsep=1pt,
        left=2pt,
        right=2pt,
        top=2pt,
        bottom=1pt,
        width=\linewidth
    ]
    \begin{verbatim}
{
    "Rule Type": "Format Compliance",
    "Rule Name": "ABV Must Follow 0.XXX Format",
    "Rule Description": "The `abv` column must be a decimal 
    between 0 and 1 with a leading zero and up to three 
    decimal places (e.g., 0.050).",
    "Target Columns": ["abv"],
    "Additional Information": {
      "Specification": "Regex ^0\\.[0-9]{2,3}$ ; numeric 
      range 0 < abv < 1.",
      "Pseudocode": [
        "if abv is null → flag",
        "if not re_match(^0\\.[0-9]{2,3}$, abv) → flag",
        "if float(abv) <= 0 or float(abv) >= 1 → flag"
      ]
    }
  }
    \end{verbatim}
    \end{tcolorbox}
     \vspace{-3mm}
    \begin{tcolorbox}[
        colback=orange!5!white,
        colframe=orange!35!white,
        title=\textcolor{black}{\textbf{Rule Card 2:}},
        boxsep=1pt,
        left=2pt,
        right=2pt,
        top=2pt,
        bottom=1pt,
        width=\linewidth
    ]
    \begin{verbatim}
  {
    "Rule Type": "Format Compliance",
    "Rule Name": "State Must Be Two-Letter Uppercase Code",
    "Rule Description": "The `state` column must hold a valid two-letter 
    U.S. state abbreviation in uppercase (e.g., \"OR\", \"IN\").",
    "Target Columns": ["state"],
    "Additional Information": {
      "Specification": "Use states.csv state list for validation.",
      "Pseudocode": [
        "if state is null → flag",
        "if len(state) != 2 → flag",
        "if state.upper() not in states.csv → flag"
      ]
    }
  }
    \end{verbatim}
    \end{tcolorbox}
 \vspace{-3mm}
    \begin{tcolorbox}[
        colback=green!5!white,
        colframe=green!35!white,
        title=\textcolor{black}{\textbf{Rule Card 3:}},
        boxsep=1pt,
        left=2pt,
        right=2pt,
        top=2pt,
        bottom=1pt,
        width=\linewidth
    ]
    \begin{verbatim}
{
    "Rule Type": "Format Compliance",
    "Rule Name": "Beer Name May Contain Letters, Numbers, Spaces",
    "Rule Description": "The `beer-name` column may include letters, 
    numbers, spaces, and apostrophes but no other special characters.",
    "Target Columns": ["beer-name"],
    "Additional Information": {
      "Specification": "Regex ^[A-Za-z0-9' ]+$ .",
      "Pseudocode": [
        "if beer_name is null → flag",
        "if not re_match(^[A-Za-z0-9' ]+$, beer_name) → flag"
      ]
    }
  }
    \end{verbatim}
    \end{tcolorbox}

    \caption{Enriched rule cards for Rule Type `Format Compliance' for Beers dataset \cite{Beers.dataset}}
    \label{fig:enriched_rule_cards_for_beer_dataset}
\end{figure}

\begin{figure}[ht]
  \tiny
  \centering
  \begin{tcolorbox}[colback=cyan!15!white,colframe=cyan!55!white,
                    title=\textcolor{black}{\textbf{Task:}},                    boxsep=1pt,left=2pt,right=2pt,top=2pt,bottom=1pt,
                    width=\linewidth]
\textit{
You are a data-quality auditor.  
Given a JSON list of rule cards, your job is to flag pairs that cannot be enforced together (same column, incompatible expectations).
}
\texttt{TASK
1. Read the rule list below.  \\
2. Detect every conflicting pair (one JSON object per pair).  \\
3. For each pair, output: \\
   * `Rule Name` – the two rule titles in their original order \\
   * `Target Columns` – the shared column names  \\
   * `Conflict Reason` – $\leq$ 20-word explanation \\  
   * `Remove Rule` – exactly one rule title to drop  \\
   * `Remove Reason` – $\leq$ 15-word justification  \\
}

\end{tcolorbox}

\vspace{-3mm}
  \begin{tcolorbox}[colback=blue!15!white,colframe=blue!45!white,
                    title=\textcolor{black}{\textbf{Do's and Dont's}},
                    boxsep=1pt,left=2pt,right=2pt,top=2pt,bottom=0pt,
                    width=\linewidth]
                    
\texttt{Formatting Do's \\
* Return a single JSON object under the key `conflicts`.  \\
* Wrap the JSON in triple back-ticks: ```json … ```.  \\
* Use the property names shown below—no extras, no re-ordering. \\
Formatting Don'ts \\
* Do not write narrative text outside the fenced JSON.   \\
* Omit pairs where `remove\_rule` would be `None`.  \\
* Report each pair once; no duplicate or transitive listings. \\
} 
\end{tcolorbox}
\vspace{-3mm}
  \begin{tcolorbox}[colback=pink!15!white,colframe=pink!45!white,
                    title=\textcolor{black}{\textbf{Few-shot examples}},
                    boxsep=1pt,left=2pt,right=2pt,top=2pt,bottom=0pt,
                    width=\linewidth]
\begin{verbatim}
    ```json
{
  "conflicts": [
    {
      "rule_names": ["Gender Must Be Binary",
                     "Gender Includes Non-Binary Options"],
      "target_column": "gender",
      "conflict_reason": "allowed value lists disagree",
      "remove_rule": "Gender Must Be Binary",
      "removal_reason": "less inclusive"
    }
  ]
}
```
\end{verbatim}                    
\end{tcolorbox}
\vspace{-3mm}
  \begin{tcolorbox}[colback=yellow!15!white,colframe=yellow!45!white,
                    title=\textcolor{black}{\textbf{Test instance}},
                    boxsep=1pt,left=2pt,right=2pt,top=2pt,bottom=0pt,
                    width=\linewidth]
\texttt{Now analyse the following rules}
\begin{verbatim}
```json
    [
  {
  "Rule Type": "Reference Table Verification",
  "Rule Name": "State Must Follow US State Code Format",
  "Rule Description": "The `state` column must contain a two-letter 
  abbreviation (e.g., NY, CO, CA, FL). Any value not on the official list 
  is invalid.",
  "Target Columns": ["state"],
  "Reference Table": "uscities.csv",
  "Additional Information": {
    "Specification": "Validate against the two-letter state_id
    field in `uscities.csv`; ignore `Country_phone_codes.csv`, 
    which is unrelated.",
    "Pseudocode": [
      "if state is null → flag",
      "if len(state) != 2 → flag",
      "if state.upper() not in us_state_abbrevs_from_csv → flag"
    ]
  }
},
{
    "Rule Type": "Format Compliance",
    "Rule Name": "ABV Must Follow 0.XXX Format",
    "Rule Description": "The `abv` column must be a decimal 
    between 0 and 1 with a leading zero and up to three 
    decimal places (e.g., 0.050).",
    "Target Columns": ["abv"],
    "Additional Information": {
      "Specification": "Regex ^0\\.[0-9]{2,3}$ ; numeric 
      range 0 < abv < 1.",
      "Pseudocode": [
        "if abv is null → flag",
        "if not re_match(^0\\.[0-9]{2,3}$, abv) → flag",
        "if float(abv) <= 0 or float(abv) >= 1 → flag"
      ]
    }
  },
  {
    "Rule Type": "Format Compliance",
    "Rule Name": "State Must Be Two-Letter Uppercase Code",
    "Rule Description": "The `state` column must hold a valid two-letter 
    U.S. state abbreviation in uppercase (e.g., \"OR\", \"IN\").",
    "Target Columns": ["state"],
    "Additional Information": {
      "Specification": "Use states.csv state list for validation.",
      "Pseudocode": [
        "if state is null → flag",
        "if len(state) != 2 → flag",
        "if state.upper() not in states.csv → flag"
      ]
    }
  },........
```
\end{verbatim}

\texttt{Return your answer in the strict JSON form described above.}

"""

\end{tcolorbox}

  \caption{Conflict resolution prompt for Beers dataset \cite{Beers.dataset}}
  \label{fig:conflict_resolution_prompt}
\end{figure}
\begin{figure}[h]
    \tiny
    \setlength{\topsep}{0pt}
    \setlength{\partopsep}{0pt}
    \setlength{\parskip}{0pt}
    \centering
    \begin{tcolorbox}[
        colback=blue!5!white,
        colframe=blue!35!white,
        title=\textcolor{black}{\textbf{Rule Card 1:}},
        boxsep=1pt,
        left=2pt,
        right=2pt,
        top=2pt,
        bottom=1pt,
        width=\linewidth
    ]
    \begin{verbatim}
  {
    "Rule Name": "ABV Must Be 0–1 Ratio",
    "Rule Description": "The `abv` column must be a decimal between 0 and 1 
    (e.g., 0.050).",
    "Target Columns": ["abv"],
    "Additional Information": {
      "Specification": "`abv` must match ^0\\.[0-9]{2,3}$ and 0 < abv < 
      1.",
      "Pseudocode": [
        "if abv is null → flag",
        "if not re_match(^0\\.[0-9]{2,3}$, abv) → flag",
        "if float(abv) <= 0 or float(abv) >= 1 → flag"
      ]
    }
  }
    \end{verbatim}
    \end{tcolorbox}
     \vspace{-3mm}
    \begin{tcolorbox}[
        colback=orange!5!white,
        colframe=orange!35!white,
        title=\textcolor{black}{\textbf{Rule Card 2:}},
        boxsep=1pt,
        left=2pt,
        right=2pt,
        top=2pt,
        bottom=1pt,
        width=\linewidth
    ]
    \begin{verbatim}
{
    "Rule Name": "ABV Must Be Percentage 0–100",
    "Rule Description": "The `abv` column must represent a percentage 
    between 0 and 100.",
    "Target Columns": ["abv"],
    "Additional Information": {
      "Specification": "`abv` must match ^\\d{1,3}(\\.\\d+)?$ and 0 < abv 
      \leq 100.",
      "Pseudocode": [
        "if abv is null → flag",
        "if not re_match(^\\d{1,3}(\\.\\d+)?$, abv) → flag",
        "if float(abv) <= 0 or float(abv) > 100 → flag"
      ]
    }
  }
    \end{verbatim}
    \end{tcolorbox}
 \vspace{-3mm}
    \begin{tcolorbox}[
        colback=pink!5!white,
        colframe=pink!35!white,
        title=\textcolor{black}{\textbf{Rule Card 3:}},
        boxsep=1pt,
        left=2pt,
        right=2pt,
        top=2pt,
        bottom=1pt,
        width=\linewidth
    ]
    \begin{verbatim}
{
    "Rule Name": "State Must Be Two-Letter Uppercase",
    "Rule Description": "The `state` column must be a valid two-letter U.S. 
    code in uppercase (e.g., OR, IN).",
    "Target Columns": ["state"],
    "Additional Information": {
      "Specification": "`state` must match ^[A-Z]{2}$ and be in 
      us_state_list.",
      "Pseudocode": [
        "if state is null → flag",
        "if not re_match(^[A-Z]{2}$, state) → flag",
        "if state not in us_state_list → flag"
      ]
    }
    \end{verbatim}
    \end{tcolorbox}
    \vspace{-3mm}
    \begin{tcolorbox}[
        colback=violet!5!white,
        colframe=violet!35!white,
        title=\textcolor{black}{\textbf{Rule Card 4:}},
        boxsep=1pt,
        left=2pt,
        right=2pt,
        top=2pt,
        bottom=1pt,
        width=\linewidth
    ]
    \begin{verbatim}
{
    "Rule Name": "State Must Be Two-Letter Lowercase",
    "Rule Description": "The `state` column must be a valid two-letter U.S.
    code in lowercase (e.g., or, in).",
    "Target Columns": ["state"],
    "Additional Information": {
      "Specification": "`state` must match ^[a-z]{2}$ and be in 
      us_state_list.",
      "Pseudocode": [
        "if state is null → flag",
        "if not re_match(^[a-z]{2}$, state) → flag",
        "if state.upper() not in us_state_list → flag"
      ]
    }
  }
    \end{verbatim}
    \end{tcolorbox}
    \caption{Generated Rule Cards which are given as input to conflict resolution module for Beers dataset \cite{Beers.dataset}}
    \label{fig:input_conflict_resolution}
\end{figure}

\begin{figure}[h]
    \tiny
    \setlength{\topsep}{0pt}
    \setlength{\partopsep}{0pt}
    \setlength{\parskip}{0pt}
    \centering
    \begin{tcolorbox}[
        colback=purple!5!white,
        colframe=purple!35!white,
        title=\textcolor{black}{\textbf{Conflict Resolve 1:}},
        boxsep=1pt,
        left=2pt,
        right=2pt,
        top=2pt,
        bottom=1pt,
        width=\linewidth
    ]
    \begin{verbatim}
 {
  "conflicts": [
    {
      "rule_names": ["ABV Must Be 0–1 Ratio", "ABV Must Be 
      Percentage 0–100"],
      "target_column": "abv",
      "conflict_reason": "ratio vs percentage scale",
      "remove_rule": "ABV Must Be Percentage 0–100",
      "removal_reason": "incompatible with sample decimals"
    }
  ]
}
    \end{verbatim}
    \end{tcolorbox}
    \vspace{-3mm}
    \begin{tcolorbox}[
        colback=green!5!white,
        colframe=green!35!white,
        title=\textcolor{black}{\textbf{Conflict Resolve 2:}},
        boxsep=1pt,
        left=2pt,
        right=2pt,
        top=2pt,
        bottom=1pt,
        width=\linewidth
    ]
    \begin{verbatim}
 {
  "conflicts": [
    {
      "rule_names": ["State Must Be Two-Letter Uppercase", 
      "State Must Be Two-Letter Lowercase"],
      "target_column": "state",
      "conflict_reason": "uppercase vs lowercase requirement",
      "remove_rule": "State Must Be Two-Letter Lowercase",
      "removal_reason": "dataset uses uppercase codes"
    }
  ]
}
    \end{verbatim}
    \end{tcolorbox}
    \caption{Output generated by the conflict resolution module}
    \label{fig:output_conflict_resolution}
\end{figure}

\begin{figure*}[h]
    \tiny
    \setlength{\topsep}{0pt}
    \setlength{\partopsep}{0pt}
    \setlength{\parskip}{0pt}
    \centering
    \begin{tcolorbox}[
        colback=cyan!15!white,
        colframe=cyan!55!white,
        title=\textcolor{black}{\textbf{Task:}},
        boxsep=1pt,
        left=2pt,
        right=2pt,
        top=2pt,
        bottom=1pt,
        width=\linewidth
    ]
    \textit{SYSTEM
    You are a senior data-quality engineer.\\
    - USER
    Your task is to audit a set of draft rules.  
    For each rule, decide which quality labels apply and justify your choice.
    } \\
    \end{tcolorbox}
    \vspace{-3mm}
    \begin{tcolorbox}[
        colback=blue!15!white,
        colframe=blue!45!white,
        title= \textcolor{black}{\textbf{Example Schema:}},
        boxsep=1pt,
        left=2pt,
        right=2pt,
        top=2pt,
        bottom=0pt,
        width=\linewidth
    ]
\textit{Label catalogue  (choose any, including none)\\
| Code | Meaning | Quick test |\\
|------|---------|------------|\\
| duplicate | Verbatim twin of another rule | Removing it changes nothing |\\
| redundant | Fully covered by a stricter rule | Violations already caught |\\
| trivial | Guaranteed by DDL / type / PK | Adds no extra protection |\\
| risk\_false\_positive | Likely to break as data evolve (hard dates, static lists, volatile policy limits) | High churn risk |\\
| mis-categorised | Belongs to a different DQ dimension | Wrong rubric |\\
| ambiguous | Unclear wording or missing comparator | Multiple readings |\\
| hallucinated\_overly\_specific | Relies on invented or unverifiable facts | No authoritative source |\\
| high\_value | Precise, stable, non-trivial and catches real errors | Keep |\\
| correct | Sound rule that needs no change but not especially high value | Keep |\\
| incorrect\_fixable | Flawed but fixable with minor edits | Revise |\\
(You may assign more than one label to a rule.)\\}
\end{tcolorbox}
    \vspace{-3mm}
     \begin{tcolorbox}[
        colback=purple!5!white,
        colframe=purple!25!white,
        title=\textcolor{black}{\textbf{Few-shot Examples:}},
        boxsep=1pt,
        left=2pt,
        right=2pt,
        top=2pt,
        bottom=0pt,
        width=\linewidth
    ]
        \begin{verbatim}
```json
[
  { "rule_name": "Order Date Must Be YYYY-MM-DD",
    "labels":    ["high_value"],
    "rationale": "precise ISO-date check adds coverage" },

  { "rule_name": "Order ID Must Be Integer",
    "labels":    ["trivial"],
    "rationale": "column is already INTEGER in schema" },

  { "rule_name": "Customer ID Must Be Positive",
    "labels":    ["duplicate"],
    "rationale": "same as 'ID Must Be Positive'" },

  { "rule_name": "Colour Must Be Red",
    "labels":    ["hallucinated_overly_specific","risk_false_positive"],
    "rationale": "no domain source; too narrow" }
]
```
    \end{verbatim}
    \end{tcolorbox}
    \vspace{-4mm}
     \begin{tcolorbox}[
        colback=orange!5!white,
        colframe=orange!25!white,
        title=\textcolor{black}{\textbf{Do's and Dont's}},
        boxsep=1pt,
        left=2pt,
        right=2pt,
        top=2pt,
        bottom=0pt,
        width=\linewidth
    ]
    \textit{Output rules\\
Dos\\
- Return a single fenced JSON array.\\
- Use keys rule\_name, labels, rationale.\\
- Keep each rationale $\leq$ 20 words.\\
Donts\\
- Do not echo the rule text or write prose outside the JSON block.\\
- Do not list a pair twice or produce transitive duplicates.
}
    \end{tcolorbox}
    
    \vspace{-3mm}
     \begin{tcolorbox}[
        colback=purple!5!white,
        colframe=purple!25!white,
        title=\textcolor{black}{\textbf{Domain Few-shot Examples}},
        boxsep=1pt,
        left=2pt,
        right=2pt,
        top=2pt,
        bottom=0pt,
        width=\linewidth
    ]
    \begin{verbatim}
    [
  { "rule_name": "Email Must Follow Standard Format",
    "labels": ["high_value"],
    "rationale": "precise regex catches common email typos" },

  { "rule_name": "Campaign ID Must Start With 'CMP' + 6 Digits",
    "labels": ["high_value"],
    "rationale": "consistent primary key across campaigns" },

  { "rule_name": "Discount Rate Must Not Exceed 20 %",
    "labels": ["risk_false_positive"],
    "rationale": "business may raise limit seasonally" },

  { "rule_name": "Customer Name Must Be at Least 3 Characters",
    "labels": ["trivial","ambiguous"],
    "rationale": "length already enforced; unclear lower bound" },

  { "rule_name": "Country Code Must Be 'US'",
    "labels": ["hallucinated_overly_specific"],
    "rationale": "marketing database contains multiple regions" }
]...
    \end{verbatim}
    \end{tcolorbox}
    
    \vspace{-4mm}
     \begin{tcolorbox}[
        colback=green!5!white,
        colframe=green!25!white,
        title=\textcolor{black}{\textbf{Input Block:}},
        boxsep=1pt,
        left=2pt,
        right=2pt,
        top=2pt,
        bottom=0pt,
        width=\linewidth
    ]
        \texttt{\\
    (A) Table schema\\
    test\_schema  \\
    (B) Candidate rules\\  
    rulelist\\
    (C) Ten-row sample\\  
    sample\_rows  
    }
    \end{tcolorbox}    
    \vspace{-3mm}
     \begin{tcolorbox}[
        colback=olive!5!white,
        colframe=olive!25!white,
        title=\textcolor{black}{\textbf{Output Format:}},
        boxsep=1pt,
        left=2pt,
        right=2pt,
        top=2pt,
        bottom=0pt,
        width=\linewidth
    ]
        \texttt{OUTPUT FORMAT\\}
    \begin{verbatim}
    [
      { "rule_name": "<exact rule name>",
        "labels":    ["duplicate", "trivial"],
        "rationale": "<= 20 words" },
      …
    ]
    \end{verbatim}
    \texttt{}
    \end{tcolorbox}
    
    \caption{Prompt Template to generate Rubric for the given Rule Cards (agnostic to Rule Type).}
    \label{fig:rubric_prompt}
\end{figure*}

\begin{figure}[h]
    \tiny
    \setlength{\topsep}{0pt}
    \setlength{\partopsep}{0pt}
    \setlength{\parskip}{0pt}
    \centering
    \begin{tcolorbox}[
        colback=cyan!15!white,
        colframe=cyan!55!white,
        title=\textcolor{black}{\textbf{Rule Card:}},
        boxsep=1pt,
        left=2pt,
        right=2pt,
        top=2pt,
        bottom=1pt,
        width=\linewidth
    ]
    \begin{verbatim}
      {
    "Rule Name": "ABV Must Be 0–1 Ratio",
    "Rule Description": "The `abv` column must be a decimal between 0 and 1 
    (e.g., 0.050).",
    "Target Column": ["abv"],
    "Additional Information": {
      "Specification": "`abv` must match ^0\\.[0-9]{2,3}$ and 0 < abv < 
      1.",
      "Pseudocode": [
        "if abv is null → flag",
        "if not re_match(^0\\.[0-9]{2,3}$, abv) → flag",
        "if float(abv) <= 0 or float(abv) >= 1 → flag"
      ]
    }
  }
    \end{verbatim}    
    \end{tcolorbox}  
    \vspace{-3mm}
    \begin{tcolorbox}[
        colback=violet!15!white,
        colframe=violet!55!white,
        title=\textcolor{black}{\textbf{Rubric Recommendation:}},
        boxsep=1pt,
        left=2pt,
        right=2pt,
        top=2pt,
        bottom=1pt,
        width=\linewidth
    ]
    \begin{verbatim}
      {
  "rule_name": "ABV Must Follow 0.XXX Decimal Format",
  "labels": ["high_value", "correct"],
  "rationale": "Precise format+range check for ABV, not covered
  elsewhere and unlikely to generate false positives."
}

    \end{verbatim}    
    \end{tcolorbox}

    \vspace{2mm}
    \begin{tcolorbox}[
        colback=green!15!white,
        colframe=green!55!white,
        title=\textcolor{black}{\textbf{Rule Card:}},
        boxsep=1pt,
        left=2pt,
        right=2pt,
        top=2pt,
        bottom=1pt,
        width=\linewidth
    ]
    \begin{verbatim}
{
    "Rule Name": "State Must Be Two-Letter Uppercase",
    "Rule Description": "The `state` column must be a valid two-letter U.S. 
    code in uppercase (e.g., OR, IN).",
    "Target Column": ["state"],
    "Additional Information": {
      "Specification": "`state` must match ^[A-Z]{2}$ and be in 
      us_state_list.",
      "Pseudocode": [
        "if state is null → flag",
        "if not re_match(^[A-Z]{2}$, state) → flag",
        "if state not in us_state_list → flag"
      ]
    }

    \end{verbatim}    
    \end{tcolorbox} 
    \vspace{-3mm}
    \begin{tcolorbox}[
        colback=olive!15!white,
        colframe=olive!55!white,
        title=\textcolor{black}{\textbf{Rubric Recommendation:}},
        boxsep=1pt,
        left=2pt,
        right=2pt,
        top=2pt,
        bottom=1pt,
        width=\linewidth
    ]
    \begin{verbatim}
      {
  "rule_name": "State Must Be Two-Letter Uppercase Code",
  "labels": ["high_value", "correct"],
  "rationale": "Standard US state abbreviation check; clear, 
  stable, not redundant with other rules."
}
    \end{verbatim}    
    \end{tcolorbox}
    \caption{Input and output rules for the rubric module.}
    \label{fig:rubric_output}
\end{figure}

\begin{figure}[ht!]
    \tiny
    \setlength{\topsep}{0pt}
    \setlength{\partopsep}{0pt}
    \setlength{\parskip}{0pt}
    \centering
    \begin{tcolorbox}[
        colback=blue!3!white,
        colframe=blue!45!white,
        title=\textcolor{black}{Code Snippet},
        boxsep=1pt,
        left=2pt,
        right=2pt,
        top=2pt,
        bottom=0pt,
        width=\linewidth
    ]
        \begin{verbatim}
import re
import pandas as pd

_ABV_PATTERN = re.compile(r"^0\.[0-9]{2,3}$")   # e.g. 0.050, 0.9 → no match

def check(df: pd.DataFrame) -> list[int]:
    """
    Flag rows whose `abv` value is null, fails the 0.xxx regex, or
    is not strictly between 0 and 1.

    Returns
    -------
    list[int]
        Sorted row indices that violate the rule.
    """
    if "abv" not in df.columns:
        raise KeyError("Missing required column: abv")

    # Treat values as strings for pattern matching
    abv_str = df["abv"].astype(str).str.strip()

    # 1. null / NaN detection (string "nan" after astype handled later)
    null_mask = df["abv"].isna()

    # 2. regex validation
    regex_mask = ~abv_str.str.match(_ABV_PATTERN)

    # 3. numeric range (convert safely)
    numeric = pd.to_numeric(df["abv"], errors="coerce")
    range_mask = (numeric <= 0) | (numeric >= 1) | numeric.isna()

    invalid_mask = null_mask | regex_mask | range_mask
    return sorted(df.index[invalid_mask].tolist())

    \end{verbatim}
    \end{tcolorbox}
    \vspace{-3mm}
    \begin{tcolorbox}[
        colback=pink!3!white,
        colframe=pink!45!white,
        title=\textcolor{black}{Code Snippet},
        boxsep=1pt,
        left=2pt,
        right=2pt,
        top=2pt,
        bottom=0pt,
        width=\linewidth
    ]
        \begin{verbatim}
import re, pandas as pd

_REGEX = re.compile(r"^[A-Z]{2}$")   # two uppercase letters

def check(df: pd.DataFrame) -> list[int]:
    if "state" not in df.columns:
        raise KeyError("state column missing")
    if "us_state_list" not in globals():
        raise NameError("define global `us_state_list`")

    s = df["state"].astype(str).str.strip().str.upper()
    bad = s.isna() | ~s.str.match(_REGEX) | ~s.isin(globals()["us_state_list"])
    return sorted(df.index[bad].tolist())

    \end{verbatim}
    \end{tcolorbox}
    \caption{Executable code snippet generated by Qwen-2.5-Coder on Beers dataset \cite{Beers.dataset}.}
    \label{fig:final_code}
\end{figure}

A multi-stage prompting strategy generates the rule cards, where each stage enriches the context provided to the language model incrementally. The process begins with a \texttt{TASK} header, which specifies the rule category, its intended scope, and the required JSON structure. Followed by are an example schema and several worked rule examples to establish a concrete pattern for the model to emulate. To adapt to the target domain, we dynamically append domain-specific few-shot examples, rules generated from previous iterations, and relevant knowledge-base snippets (e.g., ZIP codes, phone formats) to ensure alignment with the target domain and vocabulary. Finally, we add a live schema fragment from the current batch of columns (example schema is shown in Figure \ref{fig:schema_example}), keeping the prompt within the model’s context window while allowing for column-specific rule generation. An example column summary is shown in Figure \ref{fig:col_summary}.

Figure \ref{fig:rule_prompt_format_compliance} shows the complete prompt used for the \textit{Format Compliance} rule type. The initial rule cards generated for the Beers dataset \cite{Beers.dataset} appear in Figure \ref{fig:rule_cards_beers_dataset}. These are refined through an enrichment prompt (Figure \ref{fig:enrich_rule_prompt_template}), resulting in the enhanced rule cards shown in Figure \ref{fig:enriched_rule_cards_for_beer_dataset}. A conflict-resolution stage follows (prompt in Figure \ref{fig:conflict_resolution_prompt}), with input and resolved rules shown in Figure \ref{fig:input_conflict_resolution} and Figure \ref{fig:output_conflict_resolution}, respectively. Finally, we perform a rubric-based filtering using the prompt in Figure \ref{fig:rubric_prompt}, yielding the final rule cards in Figure \ref{fig:rubric_output}. The code generated corresponding to these rules is shown in Figure \ref{fig:final_code}.

Rubric-based analysis can also help assign priority levels to generated rules based on rubric recommendation labels such as `high-value`. And we can use the high-priority rules for data quality assessment in resource-constrained environments.

Appendix \ref{sec:breast_cancer_rules}, Appendix \ref{sec:bike_rules} and Appendix \ref{sec:rayyan_dataset} shows different rules generated by our system for Breast Cancer \cite{BreastCancer.dataset}, Bike \cite{Bikes.dataset} and Rayyan \cite{ouzzani2016rayyan} datasets.

\subsection{Generated Rules for the Breast-Cancer Dataset}
\label{sec:breast_cancer_rules}
\begin{table*}[ht]
\begin{tabular}{|p{5cm}|p{9cm}|}  
\hline
\textbf{Rule Type} & \textbf{Enriched Rule Card} \\ \hline

Range Constraint &
\begin{minipage}[t]{0.9\linewidth}\tiny
\begin{tcolorbox}[
        colback=orange!5!white,
        colframe=orange!35!white,
        title=\textcolor{black}{\textbf{Rule Card 1:}},
        boxsep=1pt,
        left=2pt,
        right=2pt,
        top=2pt,
        bottom=1pt,
        width=\linewidth
    ]
\begin{verbatim}
"Rule Name": "Uniformity of Cell Shape Must Be an Integer",
"Rule Description": "The `Uniformity of Cell Shape` column must 
contain only integer values.",
"Target Columns": ["Uniformity of Cell Shape"],
"Additional Information": {
    "Specification":The `Uniformity of Cell Shape` column must 
    contain only integer values between 1 and 10.",
    "Pseudocode": [
        "if Uniformity of Cell Shape is null -> flag",
        "if Uniformity of Cell Shape < 1 -> flag",
        "if Uniformity of Cell Shape > 10 -> flag"
    ]
}
\end{verbatim}
\end{tcolorbox}
\vspace{-3mm}
\begin{tcolorbox}[
        colback=pink!5!white,
        colframe=pink!35!white,
        title=\textcolor{black}{\textbf{Rule Card 2:}},
        boxsep=1pt,
        left=2pt,
        right=2pt,
        top=2pt,
        bottom=1pt,
        width=\linewidth
    ]
\begin{verbatim}
"Rule Name": "Clump Thickness Must Be an Integer",
"Rule Description": "The `Clump Thickness` column must contain 
only integer values.",
"Target Columns": ["Clump Thickness"],
"Additional Information": {
    "Specification":The `Clump Thickness` column must contain 
    only integer values between 1 and 10.",
    "Pseudocode": [
        "if Clump Thickness is null -> flag",
        "if Clump Thickness is not an integer -> flag",
        "if Clump Thickness < 1 -> flag",
        "if Clump Thickness > 10 -> flag"
    ]
}
\end{verbatim}
\end{tcolorbox}
\end{minipage}
\\ \\ \hline

Data Type Validation &
\begin{minipage}[t]{0.9\linewidth}\tiny
\begin{tcolorbox}[
        colback=blue!5!white,
        colframe=blue!15!white,
        title=\textcolor{black}{\textbf{Rule Card 1:}},
        boxsep=1pt,
        left=2pt,
        right=2pt,
        top=2pt,
        bottom=1pt,
        width=\linewidth
    ]
\begin{verbatim}
"Rule Name": "Sample Code Number Must Be an Integer",
"Rule Description": "The `Sample code number` column must contain 
only integer values.",
"Target Columns": ["Sample code number"],
"Additional Information": {
"Specification":The `Sample code number` column must contain only 
integer values.",
"Pseudocode": [
  "if sample code number is null -> flag",
  "if sample code number is not an integer -> flag"
]
}
\end{verbatim}
\end{tcolorbox}
\vspace{-3mm}
\begin{tcolorbox}[
        colback=yellow!5!white,
        colframe=yellow!15!white,
        title=\textcolor{black}{\textbf{Rule Card 2:}},
        boxsep=1pt,
        left=2pt,
        right=2pt,
        top=2pt,
        bottom=1pt,
        width=\linewidth
    ]
\begin{verbatim}
"Rule Name": "Marginal Adhesion Number Must Be an Integer",
"Rule Description": "The `Marginal Adhesion` column must contain 
only integer values.",
"Target Columns": ["Marginal Adhesion"],
"Additional Information": {
"Specification":The `Marginal Adhesion` column must contain only 
integer values.",
"Pseudocode": [
  "if marginal adhesion is null -> flag",
  "if marginal adhesion is not an integer -> flag"
]
}
\end{verbatim}
\end{tcolorbox}
\vspace{-3mm}
\begin{tcolorbox}[
        colback=magenta!5!white,
        colframe=magenta!15!white,
        title=\textcolor{black}{\textbf{Rule Card 3:}},
        boxsep=1pt,
        left=2pt,
        right=2pt,
        top=2pt,
        bottom=1pt,
        width=\linewidth
    ]
\begin{verbatim}
"Rule Name": "Single Epithelial Number Must Be an Integer",
"Rule Description": "The `Single Epithelial` column must contain 
only integer values.",
"Target Columns": ["Single Epithelial"],
"Additional Information": {
    "Specification":The `Single Epithelial` column must contain 
    only integer values.",
    "Pseudocode": [
        "if single epithelial is null -> flag",
        "if single epithelial is not an integer -> flag"
    ]
}
\end{verbatim}
\end{tcolorbox}
\vspace{-3mm}
\begin{tcolorbox}[
        colback=green!5!white,
        colframe=green!15!white,
        title=\textcolor{black}{\textbf{Rule Card 4:}},
        boxsep=1pt,
        left=2pt,
        right=2pt,
        top=2pt,
        bottom=1pt,
        width=\linewidth
    ]
\begin{verbatim}
"Rule Name": "Bare Nuclei Number Must Be an Integer",
"Rule Description": "The `Bare Nuclei` column must contain only 
integer values.",
"Target Columns": ["Bare Nuclei"],
"Additional Information": {
    "Specification":The `Bare Nuclei` column must contain only 
    integer values.",
    "Pseudocode": [
        "if bare nuclei is null -> flag",
        "if bare nuclei is not an integer -> flag"
    ]
}
\end{verbatim}
\end{tcolorbox}
\end{minipage}
\\ \hline
\end{tabular}
\caption{Rule types and their corresponding enriched rule cards for the \textit{Breast Cancer} dataset. \cite{BreastCancer.dataset}}
\label{tab:rules_generate_breast_cancer_dataset_data_type_validation}
\end{table*}

\begin{table*}[ht]
\begin{tabular}{|p{5cm}|p{9cm}|}  
\hline
\textbf{Rule Type} & \textbf{Enriched Rule Card} \\ \hline

Value Set Constraint &
\begin{minipage}[t]{0.9\linewidth}\tiny
\begin{tcolorbox}[
        colback=orange!5!white,
        colframe=orange!35!white,
        title=\textcolor{black}{\textbf{Rule Card 1:}},
        boxsep=1pt,
        left=2pt,
        right=2pt,
        top=2pt,
        bottom=1pt,
        width=\linewidth
    ]
\begin{verbatim}
"Rule Name": "Uniformity of Cell Size Must Be from Approved Set",
"Rule Description": "The `Uniformity of Cell Size` column must be 
one of [1, 4, 8, 10, 2, 3, 7, 5, 6, 9]. Any other value should be 
flagged as invalid.",
"Target Columns": ["Uniformity of Cell Size"],
"Additional Information": {
    "Specification":The `Uniformity of Cell Size` column must be 
    one of [1, 3, 5, 6, 7, 8, 9, 10].",
    "Pseudocode": [
        "if Uniformity of Cell Size is null -> flag",
        "if Uniformity of Cell Size is not in [1, 3, 5, 6, 7, 8, 
        9, 10] -> flag"
    ]
}
\end{verbatim}
\end{tcolorbox}
\vspace{-3mm}
\begin{tcolorbox}[
        colback=pink!5!white,
        colframe=pink!35!white,
        title=\textcolor{black}{\textbf{Rule Card 2:}},
        boxsep=1pt,
        left=2pt,
        right=2pt,
        top=2pt,
        bottom=1pt,
        width=\linewidth
    ]
\begin{verbatim}
"Rule Name": "Uniformity of Cell Shape Must Be from Approved Set",
"Rule Description": "The `Uniformity of Cell Shape` column must be one of [1, 4, 
8, 10, 2, 3, 5, 6, 7, 9]. Any other value should be flagged as invalid.",
"Target Columns": ["Uniformity of Cell Shape"],
"Additional Information": {
    "Specification":The `Uniformity of Cell Shape` column must be 
    one of [1, 4, 8, 10, 2, 3, 5, 6, 7, 9].",
    "Pseudocode": [
        "if Uniformity of Cell Shape is null -> flag",
        "if Uniformity of Cell Shape is not in [1, 4, 8, 10, 2, 
        3, 5, 6, 7, 9] -> flag"
    ]
}
\end{verbatim}
\end{tcolorbox}
\vspace{-3mm}
\begin{tcolorbox}[
        colback=blue!5!white,
        colframe=blue!15!white,
        title=\textcolor{black}{\textbf{Rule Card 3:}},
        boxsep=1pt,
        left=2pt,
        right=2pt,
        top=2pt,
        bottom=1pt,
        width=\linewidth
    ]
\begin{verbatim}
"Rule Name": "Marginal Adhesion Must Be from Approved Set",
"Rule Description": "The `Marginal Adhesion` column must be one 
of [1, 5, 3, 8, 10, 4, 6, 2, 9, 7]. Any other value should be 
flagged as invalid.",
"Target Columns": ["Marginal Adhesion"],
"Additional Information": {
    "Specification":The `Marginal Adhesion` column must be one of 
    [1, 5, 3, 8, 10, 4, 6, 2, 9, 7].",
    "Pseudocode": [
        "if Marginal Adhesion is null -> flag",
        "if Marginal Adhesion is not in [1, 5, 3, 8, 10, 4, 6, 2, 
        9, 7]
        -> flag"
    ]
}
\end{verbatim}
\end{tcolorbox}
\vspace{-3mm}
\begin{tcolorbox}[
        colback=yellow!5!white,
        colframe=yellow!15!white,
        title=\textcolor{black}{\textbf{Rule Card 4:}},
        boxsep=1pt,
        left=2pt,
        right=2pt,
        top=2pt,
        bottom=1pt,
        width=\linewidth
    ]
\begin{verbatim}
"Rule Name": "Single Epithelial Cell Size Must Be from Approved Set",
"Rule Description": "The `Single Epithelial Cell Size` column 
must be one of [2, 7, 3, 1, 6, 4, 5, 8, 10, 9]. Any other value 
should be flagged as invalid.",
"Target Columns": ["Single Epithelial Cell Size"],
"Additional Information": {
    "Specification":The `Single Epithelial Cell Size` column must 
    be one of [2, 7, 3, 1, 6, 4, 5, 8, 10, 9].",
    "Pseudocode": [
        "if Single Epithelial Cell Size is null -> flag",
        "if Single Epithelial Cell Size is not in [2, 7, 3, 1, 6, 
        4, 5, 8, 10, 9] -> flag"
    ]
}
\end{verbatim}
\end{tcolorbox}
\vspace{-3mm}
\begin{tcolorbox}[
        colback=magenta!5!white,
        colframe=magenta!15!white,
        title=\textcolor{black}{\textbf{Rule Card 5:}},
        boxsep=1pt,
        left=2pt,
        right=2pt,
        top=2pt,
        bottom=1pt,
        width=\linewidth
    ]
\begin{verbatim}
"Rule Name": "Bland Chromatin Must Be from Approved Set",
"Rule Description": "The `Bland Chromatin` column must be one of 
[3, 9, 1, 2, 4, 5, 7, 8, 6, 10]. Any other value should be 
flagged as invalid.",
"Target Columns": ["Bland Chromatin"],
"Additional Information": {
    "Specification":The `Bland Chromatin` column must be one of 
    [3, 9, 1, 2, 4, 5, 7, 8, 6, 10].",
    "Pseudocode": [
        "if Bland Chromatin is null -> flag",
        "if Bland Chromatin not in [3, 9, 1, 2, 4, 5, 7, 8, 6, 
        10] -> flag"
    ]
}
\end{verbatim}
\end{tcolorbox}
\vspace{-3mm}
\begin{tcolorbox}[
        colback=green!5!white,
        colframe=green!15!white,
        title=\textcolor{black}{\textbf{Rule Card 6:}},
        boxsep=1pt,
        left=2pt,
        right=2pt,
        top=2pt,
        bottom=1pt,
        width=\linewidth
    ]
\begin{verbatim}
"Rule Name": "Class Must Be from Approved Set",
"Rule Description": "The `class` column must be one of [2, 4]. 
Any other value should be flagged as invalid.",
"Target Columns": ["class"],
"Additional Information": {
    "Specification":The `class` column must be one of [2, 4]. Any 
    other value should be flagged as invalid.",
    "Pseudocode": [
      "if class is null -> flag",
      "if class is not 2 -> flag",
      "if class is not 4 -> flag"
    ]
}
\end{verbatim}
\end{tcolorbox}
\vspace{2mm}
\end{minipage}
\\ \hline
\end{tabular}
\caption{Rule types and their corresponding enriched rule cards for the \textit{Breast Cancer} dataset. \cite{BreastCancer.dataset}}
\label{tab:rules_generate_breast_cancer_dataset_value_set}
\end{table*}

\begin{table*}[ht]
\begin{tabular}{|p{5cm}|p{9cm}|}  
\hline
\textbf{Rule Type} & \textbf{Enriched Rule Card} \\ \hline

Missing Value Identification &
\begin{minipage}[t]{0.9\linewidth}\tiny
\begin{tcolorbox}[
        colback=orange!5!white,
        colframe=orange!35!white,
        title=\textcolor{black}{\textbf{Rule Card 1:}},
        boxsep=1pt,
        left=2pt,
        right=2pt,
        top=2pt,
        bottom=1pt,
        width=\linewidth
    ]
\begin{verbatim}
"Rule Name": "Sample Code Number Must Not Be NULL",
  "Rule Description": "The `Sample code number` column must 
  contain a value in every row. Any NULL or empty value should be 
  flagged as invalid.",
  "Target Columns": ["Sample code number"],
  "Additional Information": {
    "Specification":The `Sample code number` column must not be 
    null.",
    "Pseudocode": [
      "if Sample code number is null -> flag"
    ]
}
\end{verbatim}
\end{tcolorbox}
\vspace{-3mm}
\begin{tcolorbox}[
        colback=pink!5!white,
        colframe=pink!35!white,
        title=\textcolor{black}{\textbf{Rule Card 2:}},
        boxsep=1pt,
        left=2pt,
        right=2pt,
        top=2pt,
        bottom=1pt,
        width=\linewidth
    ]
\begin{verbatim}
"Rule Name": "Clump Thickness Must Be Present",
  "Rule Description": "The `Clump Thickness` column must contain 
  a value in every row. Any NULL or empty value should be flagged 
  as invalid.",
  "Target Columns": ["Clump Thickness"],
  "Additional Information": {
    "Specification":The `Clump Thickness` column must contain a 
    value in every row.",
    "Pseudocode": [
      "if Clump Thickness is null -> flag"
    ]
}
\end{verbatim}
\end{tcolorbox}
\vspace{-3mm}
\begin{tcolorbox}[
        colback=yellow!5!white,
        colframe=yellow!15!white,
        title=\textcolor{black}{\textbf{Rule Card 3:}},
        boxsep=1pt,
        left=2pt,
        right=2pt,
        top=2pt,
        bottom=1pt,
        width=\linewidth
    ]
\begin{verbatim}
"Rule Name": "Bare Nuclei Cannot Be Missing",
"Rule Description": "The `Bare Nuclei` column must not be NULL or 
empty; it must contain a value.",
"Target Columns": ["Bare Nuclei"],
"Additional Information": {
    "Specification":The `Bare Nuclei` column must not be NULL or empty",
    "Pseudocode": [
      "if Bare Nuclei is null -> flag"
    ]
}
\end{verbatim}
\end{tcolorbox}
\vspace{2mm}
\end{minipage}

\\ \hline
\end{tabular}
\caption{Rule types and their corresponding enriched rule cards for the \textit{Breast Cancer} dataset. \cite{BreastCancer.dataset}}
\label{tab:rules_generate_breast_cancer_dataset_value_presence_check}
\end{table*}

Table \ref{tab:rules_generate_breast_cancer_dataset_data_type_validation}, Table \ref{tab:rules_generate_breast_cancer_dataset_value_set}, Table \ref{tab:rules_generate_breast_cancer_dataset_value_presence_check} and Table \ref{tab:rules_generate_breast_cancer_dataset_aggregation_consistency_check} list all rules produced by our pipeline for the breast-cancer data set, grouped by rule type. Each entry reflects the enriched version of the rule—including the clarified description and pseudocode — after conflict resolution and rubric filtering.

\subsection{Generated Rules for the Bike Dataset}
 \label{sec:bike_rules}
\begin{table*}[ht]
\begin{tabular}{|p{5cm}|p{9cm}|}  
\hline
\textbf{Rule Type} & \textbf{Enriched Rule Card} \\ \hline

Pattern Matching &
\begin{minipage}[t]{0.9\linewidth}\tiny
\vspace{2mm}
\begin{tcolorbox}[
        colback=orange!8!white,
        colframe=orange!35!white,
        title=\textcolor{black}{\textbf{Rule Card 1:}},
        boxsep=1pt,
        left=2pt,
        right=2pt,
        top=2pt,
        bottom=1pt,
        width=\linewidth
    ]
\begin{verbatim}
  "Rule Name": "Article Jcreated_at Must Match Date Format",
  "Rule Description": "The `article_jcreated_at` column must match a date 
  format such as 'MM/DD/YY' or 'MM/DD/YYYY'.",
  "Target Columns": ["article_jcreated_at"],
  "Additional Information": {
    "Specification": "The `article_jcreated_at` column must follow either 
    'MM/DD/YY' or 'MM/DD/YYYY' (two- or four-digit year).",
    "Pseudocode": [
      "if article_jcreated_at is null -> flag",
      "if article_jcreated_at does not match 
      ^\\d{2}/\\d{2}/(\\d{2}|\\d{4})$ -> flag"
    ]
  }
\end{verbatim}
\end{tcolorbox}
\vspace{-3mm}
\begin{tcolorbox}[
        colback=pink!8!white,
        colframe=pink!35!white,
        title=\textcolor{black}{\textbf{Rule Card 2:}},
        boxsep=1pt,
        left=2pt,
        right=2pt,
        top=2pt,
        bottom=1pt,
        width=\linewidth
    ]
\begin{verbatim}
  "Rule Name": "Author List Must Contain Braces",
  "Rule Description": "The `author_list` column must be enclosed in curly 
  braces '{}'.",
  "Target Columns": ["author_list"],
  "Additional Information": {
    "Specification": "Each value in `author_list` must start with '{' and 
    end with '}'.",
    "Pseudocode": [
      "if author_list is null -> flag",
      "if not str(author_list).startswith('{') -> flag",
      "if not str(author_list).endswith('}') -> flag"
    ]
  }
\end{verbatim}
\end{tcolorbox}
\vspace{-3mm}
\begin{tcolorbox}[
        colback=blue!8!white,
        colframe=blue!35!white,
        title=\textcolor{black}{\textbf{Rule Card 3:}},
        boxsep=1pt,
        left=2pt,
        right=2pt,
        top=2pt,
        bottom=1pt,
        width=\linewidth
    ]
\begin{verbatim}
  "Rule Name": "Article Language Must Be Three Letters",
  "Rule Description": "The `article_language` column must be exactly 
  three alphabetic characters (e.g., 'eng', 'spa').",
  "Target Columns": ["article_language"],
  "Additional Information": {
    "Specification": "Values in `article_language` must consist of 
    exactly three letters A–Z or a–z.",
    "Pseudocode": [
      "if article_language is null -> flag",
      "if len(article_language) \notin 3 -> flag",
      "if not article_language.isalpha() -> flag"
    ]
  }
\end{verbatim}
\end{tcolorbox}
\vspace{-3mm}
\begin{tcolorbox}[
        colback=yellow!8!white,
        colframe=yellow!35!white,
        title=\textcolor{black}{\textbf{Rule Card 4:}},
        boxsep=1pt,
        left=2pt,
        right=2pt,
        top=2pt,
        bottom=1pt,
        width=\linewidth
    ]
\begin{verbatim}
  "Rule Name": "Article Language Must Be Lowercase",
  "Rule Description": "The `article_language` column must contain only 
  lowercase letters.",
  "Target Columns": ["article_language"],
  "Additional Information": {
    "Specification": "Every value in `article_language` must be 
    alphabetic and fully lowercase (e.g., 'eng', 'spa').",
    "Pseudocode": [
      "if article_language is null -> flag",
      "if not article_language.isalpha() -> flag",
      "if not article_language.islower() -> flag"
    ]
  }
\end{verbatim}
\end{tcolorbox}
\vspace{-3mm}
\begin{tcolorbox}[
        colback=pink!8!white,
        colframe=pink!35!white,
        title=\textcolor{black}{\textbf{Rule Card 5:}},
        boxsep=1pt,
        left=2pt,
        right=2pt,
        top=2pt,
        bottom=1pt,
        width=\linewidth
    ]
\begin{verbatim}
  "Rule Name": "Journal ISSN Must Match ISSN Pattern",
  "Rule Description": "The `journal_issn` column must follow the ISSN 
  format: four digits, a hyphen, and four digits (e.g., '1234-5678').",
  "Target Columns": ["journal_issn"],
  "Additional Information": {
    "Specification": "Each value in `journal_issn` must match the regex 
    ^\\d{4}-\\d{3}[\\dX]$, where the last digit may be 0–9 or 'X'.",
    "Pseudocode": [
      "if journal_issn is null -> flag",
      "if not re_match(^\\d{4}-\\d{3}[\\dX]$, journal_issn) -> flag"
    ]
  }
\end{verbatim}
\end{tcolorbox}

\vspace{-3mm}
\begin{tcolorbox}[
        colback=green!8!white,
        colframe=green!35!white,
        title=\textcolor{black}{\textbf{Rule Card 6:}},
        boxsep=1pt,
        left=2pt,
        right=2pt,
        top=2pt,
        bottom=1pt,
        width=\linewidth
    ]
\begin{verbatim}
"Rule Name": "Journal ISSN Must Have Hyphen",
  "Rule Description": "The `journal_issn` column must contain a hyphen ('-
  ') separating the first and second four-digit groups.",
  "Target Columns": ["journal_issn"],
  "Additional Information": {
    "Specification": "Each `journal_issn` value must include exactly one 
    hyphen between two character groups (e.g., '1234-5678').",
    "Pseudocode": [
      "if journal_issn is null -> flag",
      "if journal_issn.count('-') \notin 1 -> flag",
      "if '-' not in journal_issn -> flag"
    ]
  }
\end{verbatim}
\end{tcolorbox}
\vspace{2mm}
\end{minipage}
\\ \hline
\end{tabular}
\caption{Rule types and their corresponding enriched rule cards for the \textit{Rayyan} dataset. \cite{ouzzani2016rayyan}}
\label{tab:rules_generate_rayyan_dataset_pattern_matching}
\end{table*}

\begin{table*}[ht]
\begin{tabular}{|p{5cm}|p{9cm}|}  
\hline
\textbf{Rule Type} & \textbf{Enriched Rule Card} \\ \hline

Dependency Constraints &
\begin{minipage}[t]{0.9\linewidth}\tiny
\begin{tcolorbox}[
        colback=orange!8!white,
        colframe=orange!35!white,
        title=\textcolor{black}{\textbf{Rule Card 1:}},
        boxsep=1pt,
        left=2pt,
        right=2pt,
        top=2pt,
        bottom=1pt,
        width=\linewidth
    ]
\begin{verbatim}
  "Rule Name": "Journal Title Requires Journal Abbreviation",
  "Rule Description": "If `journal_title` is not null, then 
  `journal_abbreviation` must also be populated. Any journal without an 
  abbreviation is invalid.",
  "Target Columns": [
    "journal_title", "journal_abbreviation"
  ],
  "Additional Information": {
    "Specification": "Whenever `journal_title` is non-null, the companion 
    column `journal_abbreviation` must also be non-null.",
    "Pseudocode": [
      "if journal_title is null -> pass",
      "if journal_title is not null and journal_abbreviation is null -> flag"
    ]
  }
\end{verbatim}
\end{tcolorbox}
\vspace{-3mm}
\begin{tcolorbox}[
        colback=blue!8!white,
        colframe=blue!35!white,
        title=\textcolor{black}{\textbf{Rule Card 2:}},
        boxsep=1pt,
        left=2pt,
        right=2pt,
        top=2pt,
        bottom=1pt,
        width=\linewidth
    ]
\begin{verbatim}
  "Rule Name": "Journal Abbreviation Requires Journal ISSN",
  "Rule Description": "If `journal_abbreviation` is not null, then 
  `journal_issn` must also be populated. Any abbreviation without an ISSN 
  is invalid.",
  "Target Columns": [
    "journal_abbreviation", "journal_issn"
  ],
  "Additional Information": {
    "Specification": "Whenever `journal_abbreviation` has a value, the 
    column `journal_issn` must also be non-null.",
    "Pseudocode": [
      "if journal_abbreviation is null -> pass",
      "if journal_abbreviation is not null and journal_issn is null -> 
      flag"
    ]
  }
\end{verbatim}
\end{tcolorbox}
\vspace{-3mm}
\begin{tcolorbox}[
        colback=yellow!8!white,
        colframe=yellow!35!white,
        title=\textcolor{black}{\textbf{Rule Card 3:}},
        boxsep=1pt,
        left=2pt,
        right=2pt,
        top=2pt,
        bottom=1pt,
        width=\linewidth
    ]
\begin{verbatim}
  "Rule Name": "Article Volume Requires Journal Title",
  "Rule Description": "If `article_jvolumn` is not null, then `journal_title` 
  must also be populated. Any volume without a journal title is invalid.",
  "Target Columns": [
    "article_jvolumn", "journal_title"
  ],
  "Additional Information": {
    "Specification": "Whenever `article_jvolumn` contains a value, 
    `journal_title` must be non-null.",
    "Pseudocode": [
      "if article_jvolumn is null -> pass",
      "if article_jvolumn is not null and journal_title is null -> flag"
    ]
  }
\end{verbatim}
\end{tcolorbox}
\vspace{-3mm}
\begin{tcolorbox}[
        colback=pink!8!white,
        colframe=pink!35!white,
        title=\textcolor{black}{\textbf{Rule Card 4:}},
        boxsep=1pt,
        left=2pt,
        right=2pt,
        top=2pt,
        bottom=1pt,
        width=\linewidth
    ]
\begin{verbatim}
  "Rule Name": "Article Issue Requires Journal Title",
  "Rule Description": "If `article_jissue` is not null, then 
  `journal_title` must also be populated. Any issue without a journal 
  title is invalid.",
  "Target Columns": [
    "article_jissue", "journal_title"
  ],
  "Additional Information": {
    "Specification": "Whenever `article_jissue` has a value, the 
    `journal_title` column must also be non-null.",
    "Pseudocode": [
      "if article_jissue is null -> pass",
      "if article_jissue is not null and journal_title is null -> flag"
    ]
  }
\end{verbatim}
\end{tcolorbox}

\vspace{-3mm}
\begin{tcolorbox}[
        colback=green!8!white,
        colframe=green!35!white,
        title=\textcolor{black}{\textbf{Rule Card 6:}},
        boxsep=1pt,
        left=2pt,
        right=2pt,
        top=2pt,
        bottom=1pt,
        width=\linewidth
    ]
\begin{verbatim}
    "Rule Name": "Article Created Date Requires Journal Title",
    "Rule Description": "If `article_jcreated_at` is not null, then 
    `journal_title` must also be populated. Any created date without a 
    journal title is invalid.",
    "Target Columns": ["article_jcreated_at", "journal_title"],
    "Additional Information": {
      "Specification": "Whenever `article_jcreated_at` has a value, 
      `journal_title` must be non-null.",
      "Pseudocode": [
        "if article_jcreated_at is null -> pass",
        "if article_jcreated_at is not null and journal_title is null -> 
        flag"
      ]
    }
\end{verbatim}
\end{tcolorbox}
\vspace{2mm}
\end{minipage}
\\ \hline
\end{tabular}
\caption{Rule types and their corresponding enriched rule cards for the \textit{Rayyan} dataset. \cite{ouzzani2016rayyan}}
\label{tab:rules_generate_rayyan_dataset_dependency_constraint_check4}
\end{table*}

\begin{table*}[ht]
\begin{tabular}{|p{5cm}|p{9cm}|}  
\hline
\textbf{Rule Type} & \textbf{Enriched Rule Card} \\ \hline

Dependency Constraints &
\begin{minipage}[t]{0.9\linewidth}\tiny
\begin{tcolorbox}[
        colback=orange!8!white,
        colframe=orange!35!white,
        title=\textcolor{black}{\textbf{Rule Card 1:}},
        boxsep=1pt,
        left=2pt,
        right=2pt,
        top=2pt,
        bottom=1pt,
        width=\linewidth
    ]
\begin{verbatim}
"Rule Name": "Article Pagination Requires Journal Title",
"Rule Description": "If `article_pagination` is not null, then 
`journal_title` must also be populated. Any pagination without a journal 
title is invalid.",
"Target Columns": ["article_pagination", "journal_title"],
"Additional Information": {
  "Specification": "Whenever `article_pagination` has a value, 
  `journal_title` must be non-null.",
  "Pseudocode": [
    "if article_pagination is null -> pass",
    "if article_pagination is not null and journal_title is null -> flag"
  ]
}
\end{verbatim}
\end{tcolorbox}
\vspace{-3mm}
\begin{tcolorbox}[
        colback=pink!8!white,
        colframe=pink!35!white,
        title=\textcolor{black}{\textbf{Rule Card 2:}},
        boxsep=1pt,
        left=2pt,
        right=2pt,
        top=2pt,
        bottom=1pt,
        width=\linewidth
    ]
\begin{verbatim}
    "Rule Name": "Author List Requires Article Title",
    "Rule Description": "If `author_list` is not null, then `article_title` must 
    also be populated. Any author list without an article title is 
    invalid.",
    "Target Columns": ["author_list", "article_title"],
    "Additional Information": {
      "Specification": "Whenever `author_list` has a value, `article_title` must 
      be non-null.",
      "Pseudocode": [
        "if author_list is null -> pass",
        "if author_list is not null and article_title is null -> flag"
      ]
    }
\end{verbatim}
\end{tcolorbox}
\vspace{-3mm}
\begin{tcolorbox}[
        colback=blue!8!white,
        colframe=blue!35!white,
        title=\textcolor{black}{\textbf{Rule Card 3:}},
        boxsep=1pt,
        left=2pt,
        right=2pt,
        top=2pt,
        bottom=1pt,
        width=\linewidth
    ]
\begin{verbatim}
    "Rule Name": "Article Language Requires Article Title",
    "Rule Description": "If `article_language` is not null, then `article_title` 
    must also be populated. Any language without an article title is 
    invalid.",
    "Target Columns": ["article_language", "article_title"],
    "Additional Information": {
      "Specification": "Whenever `article_language` has a value, `article_title` 
      must be non-null.",
      "Pseudocode": [
        "if article_language is null -> pass",
        "if article_language is not null and article_title is null -> 
        flag"
      ]
    }
\end{verbatim}
\end{tcolorbox}
\vspace{-3mm}
\begin{tcolorbox}[
        colback=yellow!8!white,
        colframe=yellow!35!white,
        title=\textcolor{black}{\textbf{Rule Card 4:}},
        boxsep=1pt,
        left=2pt,
        right=2pt,
        top=2pt,
        bottom=1pt,
        width=\linewidth
    ]
\begin{verbatim}
    "Rule Name": "Journal Title Requires Article Title",
    "Rule Description": "If `journal_title` is not null, then `article_title` must 
    also be populated. Any journal without an article title is 
    invalid.",
    "Target Columns": ["journal_title", "article_title"],
    "Additional Information": {
      "Specification": "Whenever `journal_title` has a value, `article_title` 
      must be non-null.",
      "Pseudocode": [
        "if journal_title is null -> pass",
        "if journal_title is not null and article_title is null -> flag"
      ]
    }
\end{verbatim}
\end{tcolorbox}
\vspace{-3mm}
\begin{tcolorbox}[
        colback=pink!8!white,
        colframe=pink!35!white,
        title=\textcolor{black}{\textbf{Rule Card 5:}},
        boxsep=1pt,
        left=2pt,
        right=2pt,
        top=2pt,
        bottom=1pt,
        width=\linewidth
    ]
\begin{verbatim}
    "Rule Name": "Journal Abbreviation Requires Article Title",
    "Rule Description": "If `jounral_abbreviation` is not null, then 
    `article_title` must also be populated. Any abbreviation without an 
    article title is invalid.",
    "Target Columns": ["jounral_abbreviation", "article_title"],
    "Additional Information": {
      "Specification": "Whenever `jounral_abbreviation` has a value, 
      `article_title` must be non-null.",
      "Pseudocode": [
        "if jounral_abbreviation is null -> pass",
        "if jounral_abbreviation is not null and article_title is null 
        -> flag"
      ]
    }
\end{verbatim}
\end{tcolorbox}
\vspace{2mm}
\end{minipage}
\\ \hline
\end{tabular}
\caption{Rule types and their corresponding enriched rule cards for the \textit{Rayyan} dataset. \cite{ouzzani2016rayyan}}
\label{tab:rules_generate_rayyan_dataset_dependency_constraint_check3}
\end{table*}

\begin{table*}[ht]
\begin{tabular}{|p{5cm}|p{9cm}|}  
\hline
\textbf{Rule Type} & \textbf{Enriched Rule Card} \\ \hline

Cross Column Validation &
\begin{minipage}[t]{0.9\linewidth}\tiny
\begin{tcolorbox}[
        colback=orange!8!white,
        colframe=orange!35!white,
        title=\textcolor{black}{\textbf{Rule Card 1:}},
        boxsep=1pt,
        left=2pt,
        right=2pt,
        top=2pt,
        bottom=1pt,
        width=\linewidth
    ]
\begin{verbatim}
{
  "Rule Name": "Single Epithelial Cell Size Must Be Greater Than or Equal to 
  Mitoses",
  "Rule Description": "The value in `Single Epithelial Cell Size` must be 
  greater than or equal to the value in `Mitoses`.",
  "Target Columns": ["Single Epithelial Cell Size", "Mitoses"],
  "Additional Information": {
    "Specification": "`Single Epithelial Cell Size` must be non-null and at 
    least as large as the corresponding `Mitoses` value.",
    "Pseudocode": [
      "if Single Epithelial Cell Size is null -> flag",
      "if Mitoses is null -> flag",
      "if Single Epithelial Cell Size < Mitoses -> flag"
    ]
  }
}
\end{verbatim}
\end{tcolorbox}
\vspace{-3mm}
\begin{tcolorbox}[
        colback=pink!8!white,
        colframe=pink!35!white,
        title=\textcolor{black}{\textbf{Rule Card 2:}},
        boxsep=1pt,
        left=2pt,
        right=2pt,
        top=2pt,
        bottom=1pt,
        width=\linewidth
    ]
\begin{verbatim}
{
  
    "Rule Name": "Uniformity of Cell Shape Must Be 
    Greater Than or Equal to Clump Thickness",
    "Rule Description": "The value in the `Uniformity of Cell Shape` 
    column must be greater than or equal to the value in 
    the `Clump Thickness` column.",
    "Target Columns": ["Uniformity of Cell Shape", "Clump Thickness"],  
  "Additional Information": {
    "Specification": "`Uniformity of Cell Shape` must be non-null 
    and at least as large as the corresponding `Clump 
    Thickness` value.",
    "Pseudocode": [
      "if Uniformity of Cell Shape is null -> flag",
      "if Clump Thickness is null -> flag",
      "if Uniformity of Cell Shape < Clump Thickness -> 
      flag"
    ]
  }
}

\end{verbatim}
\end{tcolorbox}
\vspace{-3mm}
\begin{tcolorbox}[
        colback=blue!8!white,
        colframe=blue!35!white,
        title=\textcolor{black}{\textbf{Rule Card 3:}},
        boxsep=1pt,
        left=2pt,
        right=2pt,
        top=2pt,
        bottom=1pt,
        width=\linewidth
    ]
\begin{verbatim}
{
    "Rule Name": "Bland Chromatin Must Be Greater Than or 
    Equal to Clump Thickness",
    "Rule Description": "The value in `Bland Chromatin` must be 
    greater than or equal to the value in `Clump 
    Thickness`.",
    "Target Columns": ["Bland Chromatin", "Clump Thickness"],
  "Additional Information": {
    "Specification": "`Bland Chromatin` must be non-null and at 
    least as large as the corresponding `Clump Thickness` 
    value.",
    "Pseudocode": [
      "if Bland Chromatin is null -> flag",
      "if Clump Thickness is null -> flag",
      "if Bland Chromatin < Clump Thickness -> flag"
    ]
  }
}

\end{verbatim}
\end{tcolorbox}
\vspace{-3mm}
\begin{tcolorbox}[
        colback=yellow!8!white,
        colframe=yellow!35!white,
        title=\textcolor{black}{\textbf{Rule Card 4:}},
        boxsep=1pt,
        left=2pt,
        right=2pt,
        top=2pt,
        bottom=1pt,
        width=\linewidth
    ]
\begin{verbatim}
{
    "Rule Name": "Normal Nucleoli Must Be Greater Than or 
    Equal to Mitoses",
    "Rule Description": "The value in `Normal Nucleoli` must be 
    greater than or equal to the value in `Mitoses`.",
    "Target Columns": ["Normal Nucleoli", "Mitoses"],
  "Additional Information": {
    "Specification": "`Normal Nucleoli` must be non-null and at 
    least as large as the corresponding `Mitoses` value.",
    "Pseudocode": [
      "if Normal Nucleoli is null -> flag",
      "if Mitoses is null -> flag",
      "if Normal Nucleoli < Mitoses -> flag"
    ]
  }
}

\end{verbatim}
\end{tcolorbox}
\vspace{2mm}
\end{minipage}
\\ \hline
\end{tabular}
\caption{Rule types and their corresponding enriched rule cards for the \textit{Breast-Cancer} dataset. \cite{BreastCancer.dataset}}
\label{tab:rules_generate_breast_cancer_dataset_aggregation_consistency_check}
\end{table*}

Table \ref{tab:rules_generate_rayyan_dataset_pattern_matching} lists all rules produced by our pipeline for the breast-cancer data set, grouped by rule type. Each entry reflects the enriched version of the rule—including the clarified description and pseudocode — after conflict resolution and rubric filtering.

\subsection{Generated Rules for the Rayyan Dataset}
 \label{sec:rayyan_dataset}

Table \ref{tab:rules_generate_rayyan_dataset_dependency_constraint_check4} and Table \ref{tab:rules_generate_rayyan_dataset_dependency_constraint_check3} lists all rules produced by our pipeline for the Rayyan data set, grouped by rule type. Each entry reflects the enriched version of the rule—including the clarified description and pseudocode — after conflict resolution and rubric filtering.

\subsection{Results with Precision, Recall and F1 Score.}
\label{sec:precision_recall_scores}
\begin{table*}[ht]
\centering
\setlength{\tabcolsep}{1pt}            
\resizebox{\textwidth}{!}{%
\begin{tabular}{|l||
                c|c|c||c|c|c||c|c|c||c|c|c||c|c|c||c|c|c||c|c|c||c|c|c||c|c|c||}
\hline
Dataset &
\multicolumn{3}{c|}{ED2} &
\multicolumn{3}{c|}{FAHES} &
\multicolumn{3}{c|}{KATARA} &
\multicolumn{3}{c|}{IQR} &
\multicolumn{3}{c|}{IF} &
\multicolumn{3}{c|}{SD} &   
\multicolumn{3}{c|}{Max Entropy} &
\multicolumn{3}{c|}{Min-K} &
\multicolumn{3}{c|}{\textbf{Ours}} \\ \cline{2-28}
 & P & R & F$_1$ & P & R & F$_1$ & P & R & F$_1$
 & P & R & F$_1$ & P & R & F$_1$ & P & R & F$_1$
 & P & R & F$_1$ & P & R & F$_1$ & P & R & F$_1$ \\ \hline\hline
Adult &
0.75 & \textbf{0.45} & 0.57 &
0.00 & 0.00 & 0.00 &
0.01 & 1.00 & 0.02 &
0.00 & 0.00 & 0.00 &
0.00 & 0.00 & 0.00 &
0.00 & 0.00 & 0.00 &
0.79 & \textbf{0.45} & 0.57 &
0.00 & 0.00 & 0.00 &
\textbf{0.89} & 0.43 & \textbf{0.59} \\ \hline

Beers &
\textbf{1.00} & 0.97 & 0.99 &
0.73 & 0.50 & 0.59 &
0.16 & 0.56 & 0.03 &
0.00 & 0.00 & 0.00 &
0.00 & 0.00 & 0.00 &
0.00 & 0.00 & 0.00 &
0.84 & 0.97 & 0.91 &
0.91 & 0.56 & 0.69 &
\textbf{1.00} & \textbf{1.00} & \textbf{1.00} \\ \hline

Bikes &
0.58 & \textbf{0.76} & 0.65 &
0.10 & 0.23 & 0.14 &
0.19 & \textbf{0.76} & 0.30 &
0.63 & 0.17 & 0.27 &
0.83 & 0.08 & 0.14 &
0.78 & 0.13 & 0.22 &
0.63 & 0.17 & 0.27 &
0.69 & 0.20 & 0.31 &
\textbf{0.98} & 0.64 & \textbf{0.77} \\ \hline

Breast Cancer &
0.63 & 0.41 & 0.49 &   
0.06 & 0.14 & 0.09 &   
0.05 & 0.42 & 0.09 &   
0.00 & 0.00 & 0.00 &   
0.90 & 0.03 & 0.06 &   
0.00 & 0.00 & 0.00 &   
0.59 & 0.40 & 0.48 &   
0.50 & 0.20 & 0.28 &   
\textbf{0.95} & \textbf{0.72} & \textbf{0.89} \\ \hline

Flights &
 0.74 & 0.63 & 0.86 &
 0.35 & 0.01 & 0.03 &
 0.17 & 0.54 & 0.11 &
 0.00 & 0.00 & 0.00 &
 0.87 & 0.06 & 0.00 &
 0.00 & 0.00 & 0.00 &
 0.74 & \textbf{0.96} & 0.84 &
 \textbf{0.92} & 0.51 & 0.65 &
 0.75 & 0.83 & \textbf{0.89} \\ \hline

HAR &
0.67 & 0.39 & 0.48 &
0.06 & 0.14 & 0.00 &
0.04 & 0.06 & 0.05 &
0.00 & 0.00 & 0.00 &
0.91 & 0.06 & 0.11 &
0.00 & 0.00 & 0.00 &
\textbf{0.98} & \textbf{0.46} & 0.47 &
\textbf{0.98} & 0.26 & 0.41 &
\textbf{0.98} & 0.43 & \textbf{0.60} \\ \hline

Hospital &
0.83 & 0.67 & \textbf{0.99} &
0.01 & 0.03 & 0.01 &
0.05 & 0.16 & 0.08 &
0.00 & 0.00 & 0.00 &
\textbf{1.00} & 0.06 & 0.00 &
0.00 & 0.00 & 0.00 &
0.90 & \textbf{0.69} & 0.74 &
0.93 & 0.38 & 0.50 &
0.81 & 0.65 & 0.85 \\ \hline

Mercedes &
0.21 & 0.10 & 0.32 &
0.00 & 0.00 & 0.00 &
0.00 & 0.00 & 0.00 &
0.00 & 0.00 & 0.00 &
0.00 & 0.00 & 0.01 &
0.31 & 0.00 & 0.01 &
\textbf{0.42} & 0.01 & 0.21 &
0.00 & 0.00 & 0.00 &
\textbf{0.42} & \textbf{0.23} & \textbf{0.73} \\ \hline

Nasa &
0.83 & 0.73 & 0.76 &
0.08 & 0.04 & 0.05 &
0.10 & 0.17 & 0.13 &
0.00 & 0.00 & 0.00 &
1.00 & 0.06 & 0.00 &
0.00 & 0.00 & 0.00 &
0.65 & 0.80 & 0.32 &
\textbf{0.96} & 0.30 & 0.22 &
0.93 & \textbf{0.85} & \textbf{0.96} \\ \hline

Soil Moisture & 
0.60 & \textbf{0.70} & 0.05 &
0.00 & 0.00 & 0.00 &         
0.00 & 0.00 & 0.00 &         
0.03 & 0.08 & 0.00 &         
0.02 & 0.03 & 0.04 &         
0.04 & 0.02 & 0.02 &         
0.45 & 0.56 & 0.03 &         
\textbf{0.94} & 0.34 & 0.03 &
0.81 & 0.35 & \textbf{0.59} \\
\hline

\end{tabular}}
\caption{Precision(P), recall(R) and $F_{1}$ for different error detectors.}
\label{tab:rein_benchmark_full}
\end{table*}

Table \ref{tab:rein_benchmark_full} shows the earlier F1 comparison with precision-recall breakdowns, offering a fuller view of detector performance across all datasets and noise levels.

\end{document}